\documentclass[lettersize,journal]{IEEEtran}
\usepackage{amsmath,amsfonts}
\usepackage{algorithmic}
\usepackage{algorithm}
\usepackage{array}
\usepackage[caption=false,font=normalsize,labelfont=sf,textfont=sf]{subfig}
\usepackage{textcomp}
\usepackage{stfloats}
\usepackage{url}
\usepackage{verbatim}
\usepackage{graphicx}
\usepackage{cite}
\usepackage{caption}
\usepackage[numbers,sort&compress]{natbib}   
\RequirePackage{booktabs}
\usepackage{diagbox}
\usepackage{multirow}
\usepackage{amsmath,amsfonts,amssymb}
\usepackage{bm}
\usepackage{authblk}
\usepackage{xcolor}
\begin{document}

\title{\LARGE{Joint Optimization of Resource Allocation and User Association in Multi-Frequency Cellular Networks Assisted by RIS}}

\author{Yuanyuan~Qiao,
	    Yong~Niu,~\IEEEmembership{Senior Member,~IEEE}, 	
	    Zhu~Han,~\IEEEmembership{Fellow,~IEEE},
	    Shiwen~Mao,~\IEEEmembership{Fellow,~IEEE},
	    Ruisi~He,~\IEEEmembership{Senior Member,~IEEE},
      	Ning~Wang,~\IEEEmembership{Member,~IEEE}, 
	    Zhangdui~Zhong,~\IEEEmembership{Fellow,~IEEE},
	    Bo~Ai,~\IEEEmembership{Fellow,~IEEE}
    
\thanks{
	Copyright (c) 2015 IEEE. Personal use of this material is permitted. However, permission to use this material for any other purposes must be obtained from the IEEE by sending a request to pubs-permissions@ieee.org.
	This work was supported by the National Key Research and Development Program of China under Grant 2021YFB2900301; in part by the National Key Research and Development Program of China under Grant 2020YFB1806903; in part by the National Natural Science Foundation of China under Grant 62221001, Grant 62231009, Grant U21A20445; 
	This study was supported by the Fundamental Research Funds for the Central Universities, China, under grant number 2022JBQY004 and 2022JBXT001;
	in part by the Fundamental Research Funds for the Central Universities 2023JBMC030. This work is partially supported by NSF CNS-2107216, CNS-2128368, CMMI-2222810, ECCS-2302469, US Department of Transportation, Toyota and Amazon.
    The review of this article was coordinated by Xuanyu Cao. (\emph{Corresponding authors: Y. Niu, B. Ai.})
}

\thanks{Yuanyuan~Qiao is with the State Key Laboratory of Rail Traffic Control and
Safety, Beijing Jiaotong University, Beijing 100044, China, and also with
Beijing Engineering Research Center of High-speed Railway Broadband
Mobile Communications, Beijing Jiaotong University, Beijing 100044, China
(email: qiaoyuanyuan@bjtu.edu.cn).}

\thanks{Yong~Niu is with the State Key Laboratory of Rail Traffic Control and Safety,
Beijing Jiaotong University, Beijing 100044, China, and also with the National
Mobile Communications Research Laboratory, Southeast University, Nanjing
211189, China (email: niuy11@163.com).}

\thanks{Z. Han is with the Department of Electrical and Computer Engineering at the University of Houston, Houston, TX 77004 USA, and also with the Department of Computer Science and Engineering, Kyung Hee University, Seoul, South Korea, 446-701 (email: hanzhu22@gmail.com).}

\thanks{Shiwen~Mao is with the Department of Electrical and Computer Engineering,
Auburn University, Auburn, AL 36849-5201 USA (e-mail: smao@ieee.org).}

\thanks{Ruisi~He is with the State Key Laboratory of Rail Traffic Control and Safety, Beijing Jiaotong University, Beijing 100044, China, and also with Beijing Engineering Research Center of High-speed Railway Broadband Mobile Communications, Beijing Jiaotong University, Beijing 100044, China (e-mail: ruisi.he@bjtu.edu.cn).}

\thanks{Ning~Wang is with the School of Information Engineering, Zhengzhou
University, Zhengzhou, China, 450001 (email: ienwang@zzu.edu.cn).}

\thanks{Zhangdui~Zhong is with the State Key Laboratory of Rail Traffic Control
and Safety, Beijing Jiaotong University, Beijing 100044, China (e-mail:
zhdzhong@bjtu.edu.cn).}

\thanks{Bo~Ai is with the State Key Laboratory of Rail Traffic Control and
	Safety, Beijing Jiaotong University, Beijing 100044, China, and also with
	Henan Joint International Research Laboratory of Intelligent Networking
	and Data Analysis, Zhengzhou University, Zhengzhou 450001, China (email:
	boai@bjtu.edu.cn).}

}






\maketitle

\begin{abstract}
Due to the development of communication technology and the rise of user network demand, a reasonable resource allocation for wireless networks is the key to guaranteeing regular operation and improving system performance. Various frequency bands exist in the natural network environment, and heterogeneous cellular network (HCN) has become a hot topic for current research. Meanwhile, Reconfigurable Intelligent Surface (RIS) has become a key technology for developing next-generation wireless networks. By modifying the phase of the incident signal arriving at the RIS surface, RIS can improve the signal quality at the receiver and reduce co-channel interference. In this paper, we develop a RIS-assisted HCN model for a multi-base station (BS) multi-frequency network, which includes 4G, 5G, millimeter wave (mmwave), and terahertz networks, and considers the case of multiple network coverage users, which is more in line with the realistic network characteristics and the concept of 6G networks. We propose the optimization objective of maximizing the system sum rate, which is decomposed into two subproblems, i.e., the user resource allocation and the phase shift optimization problem of RIS components. Due to the NP-hard and coupling relationship, we use the block coordinate descent (BCD) method to alternately optimize the local solutions of the coalition game and the local discrete phase search algorithm to obtain the global solution. In contrast, most previous studies have used the coalition game algorithm to solve the resource allocation problem alone. Simulation results show that the algorithm performs better than the rest of the algorithms, effectively improves the system sum rate, and achieves performance close to the optimal solution of the traversal algorithm with low complexity.

\end{abstract}
\section{Introduction}

Nowadays, the global deployment of the 5G networks is gradually advancing as planned, and the 6G network is at the stage of technology research and development \cite{gaozhen1}. At the same time, thanks to the rapid development of the mobile Internet, the users' network needs are becoming more and more diverse. A single type of network cannot meet the requirements in all scenarios. Therefore, with the emergence of new networks and diversified network requirements, the emergence of heterogeneous cellular networks (HCNs) is inevitable \cite{3GPP,Lijingsurvey}. The good coverage provided by the 4G networks can meet most of the network demand. However, the throughput of the 4G networks is limited by increasingly severe interference, and scarce bandwidth \cite{4Gnoise}. The frequency of 5G is mainly divided into frequency range 1 (FR1) and FR2 \cite{5Giswhat}. The millimeter wave (mmWave) in the FR2 has the characteristics of low latency and high bandwidth because of the wavelength millimeter and the improvement of various network technologies. The mmWave networks can effectively improve the network capacity \cite{gaozhen2}, and it is being widely concerned by the research institutions \cite{5Gchallenge}. Therefore, to keep up with the explosive growth of data and the diversification of demand, it will be a significant challenge to use the current HCN fully.

Currently, widely deployed networks have different characteristics. For example, the 4G cellular network can provide good coverage and reliable data transmission. The large bandwidth in the mmWave enables it to provide high data rates \cite{5Gaibo}.
The advent of Massive MIMO has further improved the reliability and effectiveness of mmwave \cite{gaozhen3}. 
However, due to the high frequency and the short wavelength of the mmWave, it will bring about a series of problems such as the bad coverage, the serious path loss, and the high energy consumption, which are also major factors limiting the deployment of the mmWave base stations (BSs) \cite{heruisi-mmwave,zhangxiangfei-mmwave,aibommwave}. To resist the severe channel fading, the common method in the industry is to deploy the directional antenna and beamforming between the BS and the user equipment (UE) \cite{gaozhen4}.

In addition, to address the high sensitivity of mmWave communication to blocking, as well as the high dependence on the layout of the wireless environment, Reconfigurable Intelligent Surface (RIS) recently becomes an important research direction, which can effectively improve the signal quality. RIS can control the reflection of signals arriving on the RIS Surface, to enhance the received signal strength \cite{RIS1}. 
RIS is composed of multiple programmable positive-intrinsic-negative (PIN) diode elements, each of which can be used to adjust the phase shift of the signal reaching the RIS surface \cite{RIS2}. RIS can be programmed to realize the function of different phase shift adjustments of each component, to control the direct reflection of the incident signal. It can achieve the superposition of the receiver signal to improve the signal-to-interference-plus-noise ratio (SINR), and significantly improve the coverage of mmWave \cite{RIS3}.

The allocation of wireless network resources is the key that affects the performance of the BS and the user's quality of experience. It has always been a hot issue in HCN \cite{HCN,4GHCN}.
The co-channel interference is caused by spectrum sharing, which be considered in resource allocation \cite{chenyali1}. In the BS coverage area, effective resource allocation must consider resource utilization and complexity. So far, the HCN research on resource allocation has mainly focused on power control, resource allocation algorithm, and associated technology \cite{4GHCN,HCNEE,HCNLoadBalancing,HCNsinr,HCNASAPPP}. An important challenge is how to use the characteristics of HCN networks for resource allocation fully.

In this paper, we will consider the joint optimization of resource allocation and user association for HCN assisted by RIS. RIS is deployed in the mmWave BS to reduce signal interference, and improve the quality of mmWave communication. The coalition game is widely used in the nonlinear integer programming problem because of its Nash stable equilibrium \cite{chenyali2}. It is feasible to optimize the resource allocation and manage interference frequency bands. Therefore, we decompose the wireless network resource scheduling problem into two subproblems to be solved by block coordinate descent (BCD). The main contributions of this paper are as follows:

\begin{itemize}
	\item The RIS-assisted HCN model was developed using mmwave and cellular frequency types. The mmwave BS in HCN has RIS installed to provide reflected signals to improve the receiver's signal quality. We consider various cellular and mmwave network technological factors in the established HCN model. Most earlier research considered scenarios with a single BS type, frequency, or users served by a single BS. It cannot describe the complex reality of the natural environment. In contrast, the HCN model in this paper considers multiple frequency bands of 4G, 5G, mmwave, and terahertz communication networks, and users being in multiple frequency band network environments. It is more in line with the natural characteristics of complex network environments and the low, medium, and high heterogeneous network characteristics of the 6G concept \cite{6Gconcept}.
	
	\item Based on the proposed HCN model with RIS, we propose a RIS-assisted coalition game optimization algorithm by BCD to maximize the system sum rate. The coalition game algorithm is used in HCN to solve the problem of user access under multiple types of BS coverage. Meanwhile, a local search optimization method for the phase shift of RIS in mmwave BSs is proposed. Therefore, we use the BCD algorithm to solve their composite optimization problem, alternately update the local search and coalition game optimization results, and finally obtain the global solution. Due to the coupling relationship between RIS phase shift and resource allocation, this is different from most of the previous works that study the coalition game or RIS optimization singularly.
	
	\item The proposed algorithm is analyzed and compared with other schemes by configuring different network system parameters. Compared to other algorithms, the simulation results show that the algorithm optimizes user association and resource allocation to enhance the system sum rate. And comparing the proposed algorithm with the optimal solution obtained by the traversal algorithm, it can be found that the algorithm achieves a performance close to the optimal solution with low complexity.
	
\end{itemize}

The rest of this paper is organized as follows: In Section \ref{S2}, an overview of related work on the study is provided. Section \ref{S3} introduces the system model and resource allocation problem. Section \ref{S4} presents the coalition game algorithm and the local discrete phase search algorithm of RIS elements applied in the HCN model. Section \ref{S5} compares the performance differences among the proposed algorithm and several algorithms with different combinations of system parameters. Finally, Section \ref{S6} concludes this paper.

\section{Related Work}\label{S2}
At present, there have been many researches on HCN resource allocation. Coletti \emph{et al.} \cite{4GHCN} studied and compared the downlink performance of different Long Term Evolution (LTE) heterogeneous network (HetNet) deployment solutions. They focused on optimally allocating spectrum for different network layers in the evolved HetNet, including outdoor and indoor small cells. Tang \emph{et al.} \cite{HCNEE} combined the subcarrier allocation with the power allocation scheme, and developed a new resource allocation method. The method solved the quality-of-service (QoS) constrained energy-efficiency (EE) optimization in the downlink of an orthogonal frequency division multiple (OFDM) access based on two-tier HCN. Park \emph{et al.} \cite{HCNLoadBalancing} proposed a joint user association (UA) scheme with JP-CoMP using a hybrid self-organizing network (SON). The scheme is proposed for a practical clustered heterogeneous cellular network (cHCN) to maximize the network-wide proportional fairness among users. Zhou \emph{et al.} \cite{HCNsinr} proposed an HCN optimization problem with the non-orthogonal multiple access (NOMA), and designed a type of green BS assignment for NOMA-enabled HCNs. The assignment integrated with power allocation. Wei \emph{et al.} \cite{HCNASAPPP} established a general HCN model to investigate the problem of spectrum allocation to the different tiers and proposed an equivalent orthogonal network (EON) model. Su \emph{et al.} \cite{HCN5} studied the problem of resource allocation for data provided to multiple UEs in mmwave networks, and they proposed a Vickrey-Clark-Groves (VCG)-based auction mechanism to motivate UEs to disclose their evaluation of resources truthfully. Yu \emph{et al.} \cite{HCN6} proposed an intelligent driven IIoT green resource allocation mechanism for 5G heterogeneous networks, addressed by an intelligent mechanism of actor-critic driven deep reinforcement learning algorithms with asynchronous advantages. He \emph{et al.} \cite{HCN7} proposed a deep reinforcement learning-based scheme to solve the joint optimization problem of dynamic UDs-stations association and resource allocation in heterogeneous cellular network scenarios, thereby minimizing the energy consumption within a limited time delay.
The relevant studies mentioned above show that system performance can be improved by managing the resource allocation of HCNs reasonably and efficiently. However, these studies do not focus on the coverage of multi-frequency networks and multiple BSs, and the vast majority of them focus on one or a few frequency bands and a single network to cover users. In contrast to the above-related studies, this paper considers RIS-assisted HCNs composed of multiple frequency bands, including 4G, 5G, mmwave, and terahertz networks, which is consistent with the characteristics of networks in realistic environments as well as with the vision of future 6G networks \cite{6Gconcept}.


\begin{table*}[h!t]
	\renewcommand\arraystretch{1.25} 
	\centering
	\captionsetup{font={normalsize},labelsep=newline, textfont=sc}
	\caption{Differences between this paper and related works}
	\label{tab1}
	\begin{tabular}{cccccc}
		\midrule
		&
		Research scene &
		Object &
		{ Algorithm} &
		Co-Channel Interference &
		RIS \\ \hline

		this paper &
		\begin{tabular}[c]{@{}c@{}}
		Double-layer HCN \\of multi-type and multi-frequency\\ BSs(4G, 5G, mmwave, terahertz)
		\end{tabular} &
		{Max(system sum rate)} &
		\begin{tabular}[c]{@{}c@{}}
		RIS-assisted Coalition Game\\ Optimization Algorithm By BCD
		\end{tabular} &
		Co/Cross-BS &
		Yes \\

		{\cite{chenyali2}} &
		\begin{tabular}[c]{@{}c@{}}
		Double-layer HCN \\with 4G and mmwave 
		\end{tabular} &
		{ Max(system sum rate)} &
		\begin{tabular}[c]{@{}c@{}}The Coalition Formation Algorithm\\ Pairs Resource Allocation\end{tabular} &
		Co-BS &
		No \\

		{\cite{HCN5}} &
		\begin{tabular}[c]{@{}c@{}}
		Single layer HCN with mmwave
		\end{tabular} &
		Max(social-welfare) &
		\begin{tabular}[c]{@{}c@{}}
		Vickrey–Clarke–Groves (VCG) \\auction-based mechanism 
		\end{tabular} &
		Co/Cross-BS &
		No \\

		{\cite{HCN6}} &
		\begin{tabular}[c]{@{}c@{}}
		Double-layer 5G HCN\\ with single frequency 
		\end{tabular} &
		Max(energy efficiency) &
		\begin{tabular}[c]{@{}c@{}}
		Asynchronous advantage actor critic\\ driven DRL algorithm
		\end{tabular} &
		Co/Cross-BS &
		No \\	
	
		{\cite{HCN7}} &
		\begin{tabular}[c]{@{}c@{}}
		Three-layer 5G HCN\\ with double mmwave frequency 
		\end{tabular} &
		Min(energy consumption) &
		\begin{tabular}[c]{@{}c@{}}
		DRL
		\end{tabular} &
		Co/Cross-BS &
		No \\

		{\cite{HCN2}} &
		\begin{tabular}[c]{@{}c@{}}
		Double-layer HCN with mmwave 
		\end{tabular} &
		{Max(system sum rate)} &
		\begin{tabular}[c]{@{}c@{}}
		A Neural Network Approach Based \\on the encoder-decoder architecture
	   \end{tabular} &
		Co/Cross-BS &
		No \\
		
		{\cite{Coliterature2}} &
		\begin{tabular}[c]{@{}c@{}}
		Double-layer HCN with 4G
		\end{tabular} &
		\begin{tabular}[c]{@{}c@{}}
		Max(weighted\\system sum rates)
		\end{tabular} &
		\begin{tabular}[c]{@{}c@{}}
		Many-to-many Matching Game\\ Based on User Association
	   \end{tabular} &
 		Co-BS &
		No \\
		
		{\cite{Coliterature3}} &
		\begin{tabular}[c]{@{}c@{}}
		Double-layer HCN\\ with 5G low frequency
		\end{tabular} &
		\begin{tabular}[c]{@{}c@{}}
		Min(user\\ transmit power)  
		\end{tabular} &
		\begin{tabular}[c]{@{}c@{}}
		Distributive Resource Allocation\\ and User Association Algorithm
		\end{tabular} &
	    Co/Cross-BS &
		No \\

		{\cite{Coliterature4}} &
		\begin{tabular}[c]{@{}c@{}}
		Double-layer HCN \\with 4G and 5G
		\end{tabular} &
		\begin{tabular}[c]{@{}c@{}}
		Max(network-wide\\ ${\alpha }$-fair utility)
		\end{tabular} &
		The message-passing algorithm &
		Cross-BS &
		No \\
		
		{\cite{Coliterature5}} &
		\begin{tabular}[c]{@{}c@{}}
		Three-layer HCN with\\ different coverage radius of 4G
		\end{tabular} &
		\begin{tabular}[c]{@{}c@{}}
		Max(difference between\\ rate and power) 
		\end{tabular} &
		\begin{tabular}[c]{@{}c@{}}
		Multi-Agent D3QN Algorithm\\ for the Joint UARA Problem in HetNets
		\end{tabular} &
		Co-BS &
		No \\
		
		{\cite{Coliterature6}} &
		\begin{tabular}[c]{@{}c@{}}
		Multi-unit MEC network \\for heterogeneous service users
		\end{tabular} &
		Min(weighted delay) &
		The coalition game algorithm &
		Co-BS &
		No \\

		\bottomrule
	\end{tabular}
\end{table*}

Huang \emph{et al.} \cite{Coalitionnew1} proposed an exploratory coalition formation game for the uplink resource allocation problem of D2D users in hybrid networks to maximize the sum rate of the D2D system while ensuring the users' QoS requirements. Simulation results confirm that the proposed algorithm achieves near-optimal performance compared to the exhaustive algorithm and outperforms several other practical schemes in terms of the throughput of the D2D system. In literature \cite{chenyali1}, based on HCNs in sub-6 GHz and mmwave bands, a coalition formation game-based scheme is proposed to handle this challenging NP-complete optimization of the D2D resource allocation problem. Through extensive simulations with different network parameters, literature \cite{chenyali1} demonstrated the superior performance of the proposed coalition formation game compared with the optimal scheme and other sub-channel allocation schemes. Chen \emph{et al.} \cite{Coalitionnew2} investigates the cooperative reconnaissance and spectrum access (CRSA) scheme for heterogeneous federated UAV networks by joint optimization of the task layer and resource layer. It proposes a joint bandwidth allocation and coalition formation (JBACF) algorithm based on the coalition game. It verifies the effectiveness of the proposed scheme and algorithm through in-depth numerical simulations. The above-mentioned related studies show that most current studies on resource allocation in coalition games only study the application of coalition game algorithms in resource allocation separately. The RIS-assisted coalition game optimization algorithm by BCD proposed in this paper combines the coalition game and local search algorithms. It uses the BCD algorithm to optimize the two subproblems of resource allocation and the RIS phase to obtain the global solution. Compared with other algorithms, the proposed algorithm achieves good performance and near-optimal solution performance with low complexity.

To increase the throughput of enhanced mobile broadband (eMBB) users, \cite{HCN1} looks into the optimization problem of user association in URRLC mmWave communication with guaranteed reliability constraints for URLLC users. The simulation results testify to the proposed algorithm's effectiveness. \cite{HCN2} proposed a joint user association and resource allocation algorithm for a neural network-based mmwave communication system with multiple connections (MC) and integrated back propagation (IAB). First, the MIQCQP is decomposed into two subproblems, namely, the binary association matrix subproblem and the continuous IAB ratio subproblem. Based on this, they proposed a neural network to solve the binary correlation matrix inference problem and a resource allocation algorithm to find the suboptimal IAB ratio. Simulation results show that the algorithm has fast inference and good performance. \cite{HCN3} developed a new machine learning-based user association method to support multiple connections in mmwave networks. To further reduce the requirement for the number of training samples, they utilized a graphical model to represent user association scenarios and a novel feature extraction method to obtain suitable features from geolocation information and topology information. With appropriate features, each single-label classification problem can be trained in a supervised manner. Test results show that the method requires only a small number of training samples and does not require CSI to perform well.

The RIS effectively improves the quality of wireless signals with low cost and low complex structure. Therefore, the wireless communication assisted by RIS has become the key technology for developing next-generation wireless networks. There have been many studies on RIS. Liu et al. \cite{gaozhen5} proposed an elegant channel estimation solution for RIS-assisted mmWave MIMO systems, where a joint model- and data-driven paradigm is conceived. Chen et al. \cite{RISwenxian1} studied a RIS-assisted single-cell uplink communication scenario, where a cellular link and multiple D2D links share the same spectrum. Liu et al. \cite{RISwenxian2} focused on exploiting RISs in multiuser networks employing orthogonal multiple access (OMA) or NOMA. Alsenwi et al. \cite{RISwenxian3} designed algorithms to optimize the BS precoding matrix and the RIS reflecting phase shifts in a RIS-aided vehicular network. Zhu et al. \cite{RISwenxian4} considered the RIS-assisted mmWave MIMO systems, in which both the hybrid beamforming schemes at the transceivers and the RIS employ low-resolution phase shifters (PSs) for practical implementation. Wan et al. \cite{gaozhen6} proposed a novel holographic RIS, where a computationally efficient channel estimation solution and near-optimal beamforming algorithm are designed to achieve considerably improved performance. However, these works didn’t focus on resource allocation and user association assisted by RIS in HCN. 

We summarize the differences between this paper and related work in a table, as shown in Table \ref{tab1}.

\section{System Overview And Problem Formulation}\label{S3}
In this section, we first introduce the HCN model composed of multiple networks. And we describe the problem of resource allocation and user association in the HCN by expounding the calculation function of system performance, and the reflected channel of RIS in subsection \ref{S3-2}. Finally, we model the optimization problem in subsection \ref{S3-3}.

This work analyzes the RIS-assisted HCN scenario with several BSs and frequencies, in contrast to most prior research, concentrating on the resource allocation problem for a single BS or frequency. Compared to earlier work, this scenario is more complex. It requires solving user access and co-channel interference of co-BS and cross-BS, under the coverage of multiple BS.

\subsection{Description}\label{S3-1}
We built a scenario with multiple BSs and multiple UEs establishing network connections. In this paper, we study which BS the user chooses to establish the connection assisted by RIS, and achieve a balance between the interference and the system sum rate. The interference is mainly caused by users sharing the same spectrum on the same type of network.
The relative position between UEs and BSs determines the group of networks users can select. When a user is within the BS coverage area, the BS can establish a network connection with the user.

\begin{figure}[ht]
	\begin{center}
		\includegraphics[height=3.25in]{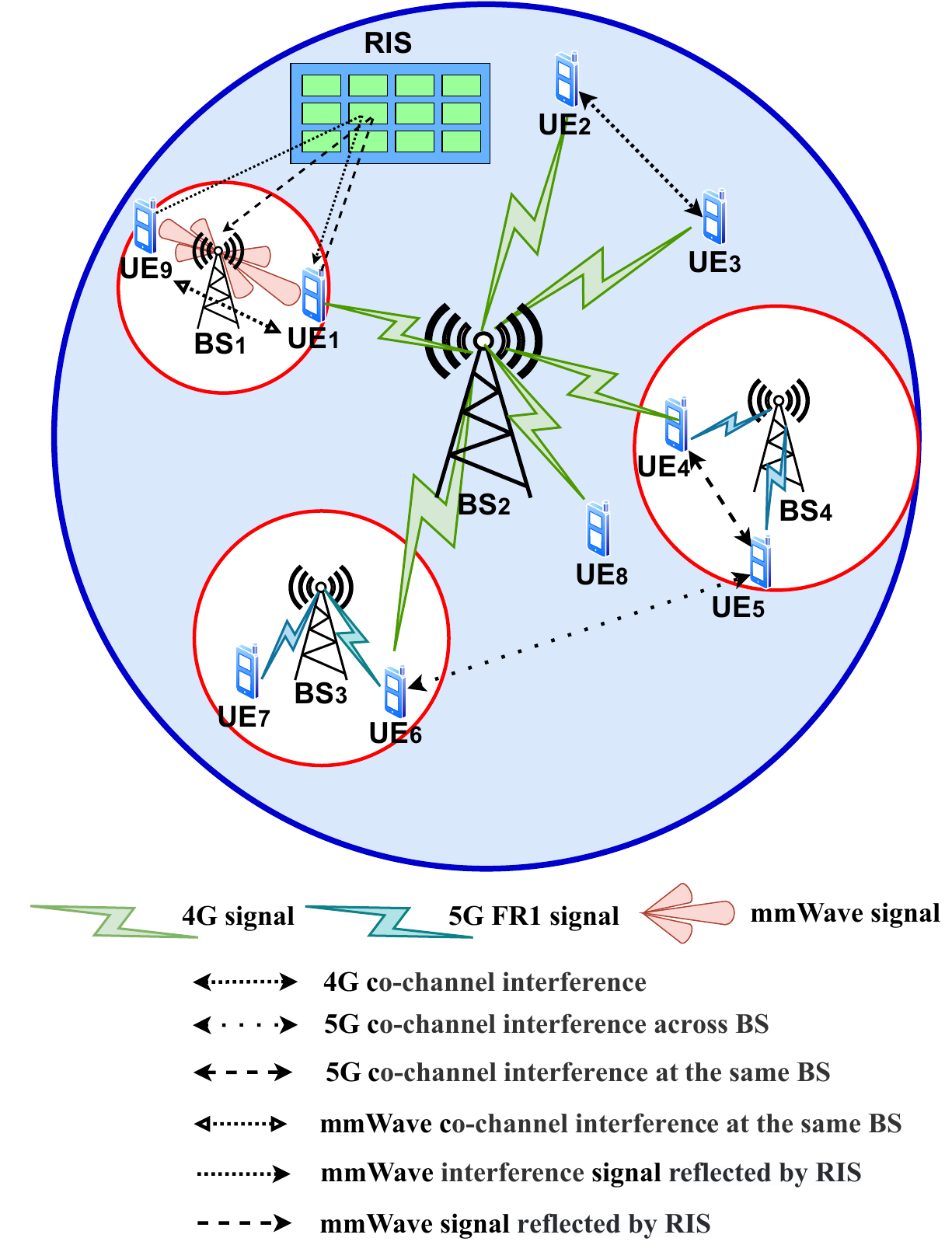}
	\end{center}
	\caption{System model for a heterogeneous wireless network of the user assocation assisted by RIS, where there are 4 BSs and 9 users.} \label{fig1}
\end{figure}

As shown in Fig. \ref{fig1}, there are 4 BSs and 9 UEs, that the ${BS_{1}}$ is a mmWave BS, the ${BS_{2}}$ is a 4G BS, the ${BS_{3}}$ and the ${BS_{4}}$ are both 5G BSs. The mmWave BS is equipped with the RIS. The RIS is composed of programmable elements, each of which individually reflects the signal reaching the RIS surface and transmits it to the receiver to enhance the received signal. Therefore, the mmWave BS can receive direct signals from the user and reflected signals assisted by RIS. The signal reflected by RIS interferes only between users occupying the same spectrum.
Whether a UE can establish a connection with a BS depends on whether the UE is within the BS coverage area and interference caused by the shared spectrum.
In this paper, when we consider the mmWave networks, it is not the same as other types of networks in terms of antenna construction. The mmWave communication is equipped with the high-directional antenna for directional transmission of high-gain communication among BSs and UEs to resist the severe path loss of the mmWave \cite{Niummwave}. In this paper, interference is an important factor affecting the allocation of wireless network resources.

We focus on properly managing and limiting interference for system performance in such a system. Therefore, when the users in a BS degrade the system performance due to severe co-channel interference, we can switch the users to access other BSs to reduce the same spectrum sharing. At the same time, we optimize the phase shift settings of RIS components installed in mmWave BSs to enhance the mmWave signal quality further and reduce the signal loss caused by high frequency. In the way, on the one hand, the interference among users accessed to the same-BS can be alleviated, and on the other hand, the cross-BS interference among users can also be alleviated. The model in this paper assumes that the various mmWave BSs have different frequencies and that there won't be any cross-BS interference because of the wide mmWave bandwidth.

\subsection{System Model}\label{S3-2}
In our model, we assume that a cell has ${t}$ types of BSs, including 4G, 5G, mmwave, etc. The arrangement of multiple BSs will result in a cross-coverage area of several networks, in which users in this area can choose to access these networks. These networks is called an overlapping network set, which is denoted as ${w_i}$, ${w_i} \in W$.
The ${ {i_{th}}}$ network in the overlapping network set is denoted as ${ {n_i}}$, and the ${ {i_{th}}}$ user in the coverage area is denoted as ${c_i}$. The ${ {j_{th}}}$ sub-channel in the network ${{n_i}}$ accessed by user ${c_i}$ is denoted as ${S_{{n_i},{c_i}}^j}$, ${S_{{n_i},{c_i}}^j \in S}$. The bandwidth of the sub-channel is denoted as ${B_{{n_i}}^j}$.

RIS is composed of ${N \times N}$ programmable elements PIN, each element can independently adjust the phase shift of incident signal through ON/OFF state PIN in real time, so as to enhance the signal strength at the receiver. We set the phase shifts of the elements to discrete, equally spaced finite values of ${[0,2\pi )}$.
The number of quantized bits is ${e}$, so there are ${{2^e}}$ phase shifts. We set the response coefficient of ${{l_x}}$-th row and ${{l_z}}$-th column in RIS as ${{q_{{l_x},{l_z}}} = {e^{j{\theta _{{l_x},{l_z}}}}}}$ , where the phase shift is ${{\theta _{{l_x},{l_z}}} = \frac{{2\pi {m_{{l_x},{l_z}}}}}{{{2^e} - 1}}}$, ${{m_{{l_x},{l_z}}} = \{ 0,1,...,{2^e} - 1\} }$, ${1 \le {l_x},{l_z} \le N}$ and ${j}$ represents the imaginary part.
Based on the model established in the previous section, we can find that there are two kinds of channels in mmWave communication, one is the direct channel of mmWave, and the other is the channel reflected by RIS element ${\{ {l_x},{l_z}\} }$. The channel reflected by RIS element ${\{ {l_x},{l_z}\} }$ from the transmitter ${t}$ to the receiver ${r}$ is denoted as  ${h_{{l_x},{l_z}}^{t,r}}$. In order to simply represent the reflected channel of RIS element, the combination of reflected channel and response coefficient of the RIS element is expressed as
\begin{equation}
{H_{{l_x},{l_z}}} = h_{{l_x},{l_z}}^{t,r}{q_{{l_x},{l_z}}}
\label{eq1}
\end{equation}

We define a binary variable ${ X_{{c_i}}^{{n_i}}}$ , which means that, if user ${c_i}$ is connected to network ${n_i}$ then ${ X_{{c_i}}^{{n_i}} = 1}$, otherwise ${ X_{{c_i}}^{{n_i}} = 0}$. Therefore, ${ \sum\limits_{{n_i} \in N} {X_{{c_i}}^{{n_i}} = 1} }$ means that user ${c_i}$ has and is connected to only one network. Conversely, if ${ \sum\limits_{{n_i} \in N} {X_{{c_i}}^{{n_i}} = 0} }$, then it means that user ${c_i}$ has no access to any network. 
We define a binary variable ${ X_{{c_i}}^{{c_j}}}$, indicating that for users ${c_i}$ and ${c_j}$, if both access sub-channel in the same frequency, it is noted as ${X_{{c_i}}^{{c_j}} = 1}$, otherwise ${X_{{c_i}}^{{c_j}} = 0}$.

To facilitate analysis and processing, it is assumed that networks occupied by different frequencies do not interfere with each other. Within the same BSs, different sub-channels are independent of each other and do not cause interference. Therefore, the interference comes from users that access the same type of network that share the same sub-channel.

To maximize system performance in terms of the system sum rate, we first need to consider the SINR. For the small-scale fading, we consider the rayleigh channel model \cite{Rayleigh} in direct channel. We assume that different networks have different channel models, and the power or second-order statistics of the channel is represented by the ${\left |h_0  \right |_{t}^{2} }$, which is a constant in the BS coverage area. For ease of presentation,  ${\left |h_0  \right |_{t}^{2} }$ is denoted as ${A_t}$, where ${h_0}$ is a complex Gaussian random variable with variance 1 and mean 0. For the communication connection link ${i}$, we denote its sender and receiver by ${s_i}$ and ${r_i}$, respectively. Based on the path loss model, the received power expression of ${s_i}$ at ${r_i}$ is expressed as: ${ P_{r}^{t}(i, i)=A_{t} \cdot G_{t}^{t} \cdot G_{r}{ }^{t} \cdot l_{i i}^{-u_{t}} \cdot P_{t}}$, where ${P_t}$ is the transmit power of the ${t}$th type network, ${l_{ii}}$ is the distance between ${s_i}$ and ${r_i}$, ${u_t}$ is the path loss index of the ${t}$th type network, and ${G_{t}^{t} }$ is transmit antenna gain of the ${t}$th type network, ${G_{r}^{t} }$ is the receive antenna gain of the ${t}$th type network. They are all network constants of the ${t}$th type network. ${l_{ii}^{-u_t} }$ is the path loss model, defined as ${L(d)=L\left(d_{0}\right)+10 n \lg \left(\frac{d}{d_{0}}\right)+X_{\sigma}}$. ${L(d)}$ is the free space loss, which is defined as $L\left(d_{0}\right)=32.45+20 \lg (f)+20 \lg \left(d_{0}\right)$, that ${f}$ is the signal transmission frequency, ${d_0}$ is the reference distance, and ${d}$ is the communication transmission distance. ${X_\sigma }$ is shadow fading, defined as a normal random variable with mean 0 and standard deviation ${\sigma }$ \cite{goldsmith}.

\begin{figure}[t]
	\begin{center}
		\includegraphics[height=2 in]{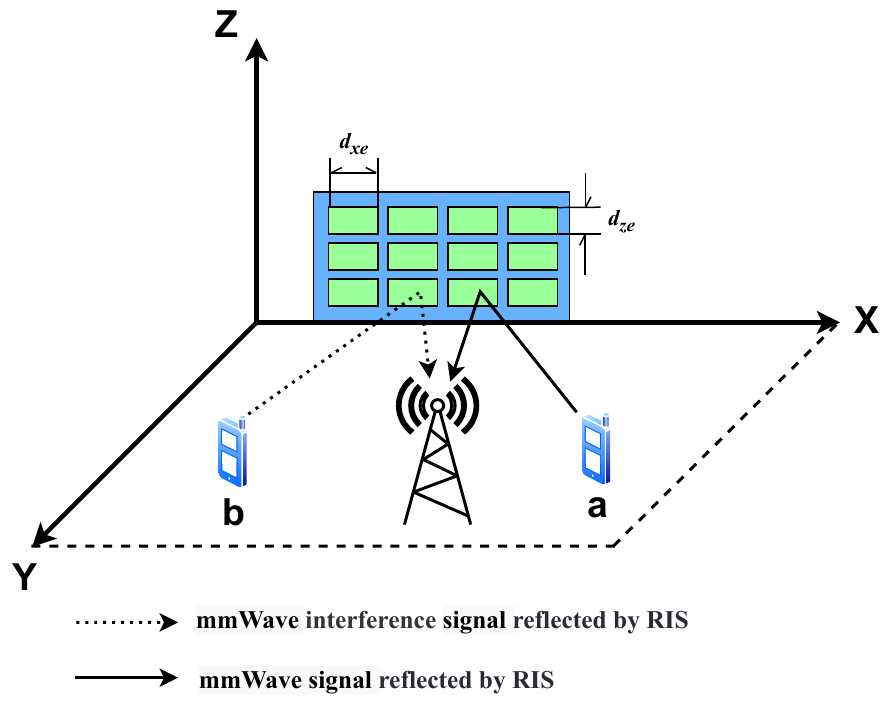}
	\end{center}
	\caption{Schematic diagram of RIS reflection signal} \label{fig2}
\end{figure}

In the mmWave communication, the RIS reflection channel model depends on the position of RIS, sender and receiver, so we first need to establish a three-dimensional Cartesian coordinate system to represent the relative position of devices in RIS channel transmission.
As shown in Fig. \ref{fig2}, RIS is set in the north direction of the mmWave BS, where the communication coverage radius of the BS is located. UE ${a}$ is the transmitter, and UE ${b}$ is the co-frequency interference device of UE ${a}$.
The coordinate of the mmWave BS is ${({R_x},{R_y},0)}$, and the communication coverage radius is ${r}$.
The distance between adjacent elements in RIS, ${{d_{xe}}}$ on the ${X}$-axis and ${{d_{ze}}}$ on the ${Z}$-axis, is 0.005 m. The coordinates of RIS elements that are horizontal and perpendicular to the mmWave BS are ${({R_x},{R_y} - r,{d_{ze}})}$, and for each RIS element coordinate, the upper right corner is recorded as its position coordinate.
Therefore, the coordinates ${({S_x},{S_y},{S_z})}$ of the RIS element ${\{ {l_x},{l_z}\} }$ can be expressed as ${({R_x} - {d_{xe}}*(N/2 + 1) + {d_{xe}}*{l_x},{R_y} - r,{d_{ze}}*{l_z})}$.

 Take an RIS reflection link ${i}$ as an example, we set the coordinates of transmitter ${{t_i}}$ and receiver ${{r_i}}$ as ${({t_{ix}},{t_{iy}},0)}$ and ${({r_{ix}},{r_{iy}},0)}$, respectively.
Therefore, in conjunction with the RIS settings, we define the distance between ${{t_i}}$ and the RIS element ${\{ {l_x},{l_z}\} }$ as ${D_{{t_i}}^{\{ {l_x},{l_z}\} }}$. The distance from ${{r_i}}$ to RIS element ${\{ {l_x},{l_z}\} }$ is defined as ${D_{{r_i}}^{\{ {l_x},{l_z}\} }}$. They can be expressed as
\begin{equation}
D_{{t_i}}^{{l_x},{l_z}} = \sqrt {{{({t_{ix}} - {S_x})}^2} + {{({t_{iy}} - {S_y})}^2} + {{( - {S_z})}^2}} 
\label{eq2},
\end{equation}
\begin{equation}
D_{{r_i}}^{{l_x},{l_z}} = \sqrt {{{({r_{ix}} - {S_x})}^2} + {{({r_{iy}} - {S_y})}^2} + {{( - {S_z})}^2}} 
\label{eq3}.
\end{equation}

Based on the constructed coordinate system, we use the Rician channel to describe the RIS reflection channel, including LoS (Line of Sight) and NLoS (None Line of Sight). Therefore, the mmWave RIS reflection channel model can be expressed as

\begin{equation}
h_{{l_x},{l_z}}^{{r_i},{t_i}} = \sqrt {\frac{\beta }{{1 + \beta }}} \tilde h_{{l_x},{l_z}}^{{r_i},{t_i}} + \sqrt {\frac{1}{{1 + \beta }}} \hat h_{{l_x},{l_z}}^{{r_i},{t_i}},
\label{eq4}
\end{equation}
where ${\beta  = 4}$ is the Rician factor, ${\tilde h_{{l_x},{l_z}}^{{r_i},{t_i}}}$ is the LoS part of RIS reflection channel, and ${\hat h_{{l_x},{l_z}}^{{r_i},{t_i}}}$ is the NLoS part of RIS reflection channel, which can be expressed as
\begin{equation}
\tilde h_{{l_x},{l_z}}^{{r_i},{t_i}} = \sqrt {{{(D_{{t_i}}^{{l_x},{l_z}} \cdot D_{{r_i}}^{{l_x},{l_z}})}^{ - \alpha }}} {e^{ - j\theta '}},
\label{eq5}
\end{equation}

\begin{equation}
\hat h_{{l_x},{l_z}}^{{r_i},{t_i}} = \sqrt {{{(D_{{t_i}}^{{l_x},{l_z}} \cdot D_{{r_i}}^{{l_x},{l_z}})}^{ - \alpha '}}} \hat h_{NLoS,{l_x},{l_z}}^{{r_i},{t_i}},
\label{eq6}
\end{equation}
where ${\theta '}$ is the phase shift valued in ${[0,2\pi ]}$, ${\alpha }$ is the path-loss index of LoS, ${\alpha '}$ is the path-loss index of NLoS, and ${\hat{h}_{NLoS, l_{z}, l_{y}}^{r_{i}, t_{i}} \sim \mathcal{C N}(0,1)}$ is the small-scale fading.

For the cellular network, the SINR of a user ${c_i}$ accessing the network can be expressed as

\begin{equation}
SINR_{{c_i}}^{cell} = \frac{{{A_t} \cdot {G_t}^t \cdot {G_r}^t \cdot l_{{n_i},{c_i}}^{ - {u_t}} \cdot {P_t}^{cell}}}{{P_{{\mathop{\rm int}} ,{c_i}}^{cell,t} + N_0^tB_{{n_i}}^j}},
\label{eq7}
\end{equation}
where ${N_0^t}$ is the single-sided white noise power spectral density of the ${t}$th type network, that ${t}$th type network is the kind of user ${c_i}$ access network. ${B_{{n_i}}^j}$ is the sub-channel bandwidth allocated by user ${c_i}$, and ${{P_t}^{cell}}$ is the transmitting power of user ${c_i}$. The interference signal power ${P_{{\mathop{\rm int}} ,{c_i}}^{cell,t}}$ received by cellular user ${c_i}$ is expressed as (8).
Therefore, according to the Shannon's theory, the uplink throughput of the cellular user ${c_i}$ can be expressed as ${R_{cell}^{{c_i}}}$, as shown in (9). 

 \begin{equation}
P_{{\rm{int}},{c_i}}^{cell,t} = \sum\limits_{z \in W} {\sum\limits_{{c_j} \in {C_z},{c_j} \ne {c_i}} {X_{{c_i}}^{{c_j}} \cdot {A_t} \cdot {G_t}^t \cdot {G_r}^t \cdot l_{{n_i},{c_j}}^{ - {u_t}} \cdot {P_t}^{cell}} } .
\label{eq8}
\end{equation}

\begin{equation}
\begin{split}
R_{cell}^{{c_i}} = \mathop \sum \limits_{{n_i} \in N} X_{{c_i}}^{{n_i}}B_{{n_i}}^j{\log _2}\left( {1 + \frac{{{A_t} \cdot G_t^t \cdot G_r^t \cdot l_{{n_i},{c_i}}^{ - {u_t}} \cdot P_t^{cell}}}{{P_{{\mathop{\rm int}} ,{c_i}}^{cell,t} + N_0^tB_{{n_i}}^j}}} \right).
\end{split}
\label{eq9}
\end{equation}

For the mmWave network, the SINR of a user ${c_i}$ accessing the network can be expressed as (10), where ${k_0}$ is a constant coefficient and is proportional to ${{\left( {\lambda /4\pi } \right)^2}}$, that ${\lambda }$ refers to the wavelength. ${{F_m} = \sqrt {l_{{n_i},{c_i}}^{ - {u_t}}}  + \sum\limits_{{l_x},{l_z}} {{H_{{l_x},{l_z}}}} }$ denotes the channel coefficient of the mmwave, which consists of the direct and reflected channels.
${P_t^m}$ is the transmit power when the mmWave user accesses to the ${t}$th type network, and the interference signal power ${P_{int,{c_i}}^{m,t} }$ received by the mmWave user ${c_i}$ is expressed as (11), where ${\rho }$ is the interference factor between different links in the network. ${P_{RIS,{c_i}}^{m,t}}$ is the interference caused by RIS reflection of mmWave, which can be expressed as (12).

 \begin{equation}
SINR_{{c_i}}^m = \frac{{{{\left| {{F_m}} \right|}^2} \cdot {k_0} \cdot {G_t}^t({n_i},{c_i}) \cdot {G_r}^t({n_i},{c_i}) \cdot P_t^m}}{{P_{RIS,{c_i}}^{m,t} + P_{{\rm{int}},{c_i}}^{m,t} + N_0^tB_{{n_i}}^j}}.
\label{eq10}
\end{equation}	

\begin{equation}
\begin{array}{*{20}{l}}
{P_{int,{c_i}}^{m,t} = \sum\limits_{z \in W} {\sum\limits_{{c_j} \in {C_z}}^{{c_j} \ne {c_i}} {X_{{c_i}}^{{c_j}} \cdot \rho  \cdot {k_0} \cdot {G_t}^t({n_i},{c_j})} } }\\
{\begin{array}{*{20}{c}}
	{}&{\begin{array}{*{20}{c}}
		{}&{}
		\end{array}}&{ \cdot {G_r}^t({n_i},{c_j}) \cdot l_{{n_i},{c_j}}^{ - {u_t}} \cdot P_t^m}
	\end{array}}
\end{array}.
\label{eq11}
\end{equation}

\begin{equation}
P_{RIS,{c_i}}^{m,t} = \sum\limits_{{c_j} \in {C_z}}^{{c_j} \ne {c_i}} {{{\left| {\sum\limits_{{l_x},{l_z}} {{H_{{l_x},{l_z}}}} } \right|}^2}}  \cdot {G_t}^t({n_i},{c_j}) \cdot {G_r}^t({n_i},{c_j}) \cdot P_t^m.
\label{eq12}
\end{equation}	

\begin{equation}
\begin{array}{l}
R_m^{{c_i}} = \sum\limits_{{n_i} \in N} {X_{{c_i}}^{{n_i}}} (1 - P_{{n_i},{c_i}}^{out})B_{{n_i}}^j\\
\begin{array}{*{20}{c}}
{}&{ \cdot {{\log }_2}\left( {1 + \frac{{{{\left| {{F_m}} \right|}^2} \cdot {k_0} \cdot {G_t}^t({n_i},{c_i}) \cdot {G_r}^t({n_i},{c_i}) \cdot P_t^m}}{{P_{RIS,{c_i}}^{m,t} + P_{{\mathop{\rm int}} ,{c_i}}^{m,t} + N_0^tB_{{n_i}}^j}}} \right)}
\end{array}
\end{array}.
\label{eq13}
\end{equation}

Therefore, the uplink throughput of the mmWave user ${c_i}$ can be expressed as ${R_m^{{c_i}}}$, as shown in (13), where ${P_{{n_i},{c_i}}^{out}}$ represents the interruption probability between the transmitter and the receiver during the mmWave transmission. And it can be denoted as ${P_{a,b}^{out} = 1 - {e^{ - \beta {l_{ab}}}}}$, where ${l_{ab}}$ is the distance between the transmitter ${a}$ and the receiver ${b}$, and ${\beta }$ is a parameter that reflects the density and size of obstacles. The obstacles may interrupt the mmWave communications due to occlusion effects \cite{Pout}.

\begin{equation}	    
{R^{{c_i}}} = \left\{ {\begin{array}{*{20}{c}}
	{R_{cell}^{{c_i}}{\kern 1pt} {\kern 1pt} {\kern 1pt} {\kern 1pt} {\kern 1pt} {\kern 1pt} {\kern 1pt} cellular{\kern 1pt} {\kern 1pt} {\kern 1pt} {\kern 1pt} user{\kern 1pt} {\kern 1pt} {\kern 1pt} c}\\
	{{\kern 1pt} {\kern 1pt} {\kern 1pt} {\kern 1pt} R_m^{{c_i}}{\kern 1pt} {\kern 1pt} {\kern 1pt} {\kern 1pt} {\kern 1pt} {\kern 1pt} {\kern 1pt} {\kern 1pt} {\kern 1pt} {\kern 1pt} {\kern 1pt} mmwave{\kern 1pt} {\kern 1pt} {\kern 1pt} {\kern 1pt} user{\kern 1pt} {\kern 1pt} {\kern 1pt} c}
	\end{array}} \right..
\label{eq14}
\end{equation}

Therefore, the uplink throughput of the user ${c_i}$, denoted as ${{R^{{c_i}}}}$, is give in (14). The sum uplink throughput of all users in the system can be expressed as (15).

\begin{equation}
R = \sum\limits_W {\sum\limits_{{c_i} \in C} {{R^{{c_i}}}} } .
\label{eq15}
\end{equation}

\subsection{Problem Formulation}\label{S3-3}
Combined with the discussion in the Section \ref{S3-2}, we model the optimization problem of resource allocation and user association assisted by RIS in the HCN model as (16).

\begin{equation}
\begin{split}
&\max_{(S,X),\Theta}{\rm{R(S,X,}}\Theta {\rm{)}} = \sum\limits_W {\sum\limits_{{c_i} \in C} {{R^{{c_i}}}} } \\
&\rm {s.t.} \\
(a)&{X_{{c_i}}^{{n_i}}} \in \{ 0,1\} ,  \forall {n_i},{c_i}, \\
(b)&\sum\limits_{{n_i} \in N} {X_{{c_i}}^{{n_i}}}  \in \left\{ {0,1} \right\}   ,\forall {n_i},{c_i},\\
(c)&{X_{{c_i}}^{{c_j}}} \in \left\{ {0,1} \right\},\forall {n_i},{c_i},\\
(d)&{B_{{n_i}}^j} \ge 0, \forall {n_i},{c_i},\\
(e)&{q_{{l_x},{l_z}}} = {e^{j{\theta _{{l_x},{l_z}}}}},{\theta _{{l_x},{l_z}}} = \frac{{2\pi {m_{{l_x},{l_z}}}}}{{{2^e} - 1}},\\
&   {m_{{l_x},{l_z}}} = \{ 0,1,...,{2^e} - 1\} ,1 \le {l_x},{l_z} \le N.
\end{split}
\label{eq16}
\end{equation}

By analyzing (14), it can be found that the system sum uplink throughput assisted by RIS is related to the sub-channel ${S_{{n_i},{c_i}}^j}$ accessed by the user ${c_i}$, the resource sharing relationship ${X_{{c_i}}^{{n_i}}}$, and phase shift ${\Theta }$ of all components of RIS. So we define the system utility function as the system sum rate, denoted as ${{\rm{R(S,X,}}\Theta {\rm{)}}}$, where ${S}$ represents the set of ${S_{{n_i},{c_i}}^j}$, ${X}$ represents the set of ${X_{{c_i}}^{{n_i}}}$, and ${\Theta  = \{ {\theta _{{l_x},{l_z}}},1 \le {l_x},{l_z} \le N\} }$. Based on the above analysis, the problem of resource allocation and user association in HCN assisted by RIS for the maximum system sum rate is denoted as (16), where (16a) indicates that if the user ${c_i}$ accesses to the network ${n_i}$, it is denoted as ${X_{{c_i}}^{{n_i}} = 1}$, otherwise ${X_{{c_i}}^{{n_i}} = 0}$. In (16b), ${\sum\limits_{{n_i} \in N} {X_{{c_i}}^{{n_i}}}  = 1}$ indicates that the user ${c_i}$ only accesses to one network. Otherwise, the user ${c_i}$ has no access to any network. (16c) indicates whether the spectrum resources occupied by the users ${c_i}$ and the user ${c_j}$ are the same. If they are the same, then ${X_{{c_i}}^{{c_j}} = 1}$, otherwise ${X_{{c_i}}^{{c_j}} = 0}$. (16d) indicates that for the user ${c_i}$, the ${B_{{n_i}}^j}$ is always greater than or equal to 0. The phase shift of (16e) is a discrete variable. It is shown that the optimization problem is NP-complete, and our goal is to obtain an allocation scheme for the maximum system sum rate.

\section{Problem Formulation}\label{S4}
Based on the system model and objective problem established in Section \ref{S3}, we divide the optimization problem into two sub-problems for the joint optimization of user association and resource allocation and the RIS phase in subsection \ref{S4-1}. We solve these two sub-problems by BCD until the convergence condition of the algorithm is met and the suboptimal solution is obtained in subsection \ref{S4-3}.

\subsection{Problem Decomposition}\label{S4-1}
To solve the sum-rate maximization problem proposed in Section \ref{S3}, we decompose the optimization problem (16) into two sub-problems, i.e., the joint optimization of user association and resource allocation, and the phase-shift optimization of RIS.

\emph{1) The Joint Optimization of User Association and Resource Allocation:} With the RIS phase fixed and constant, the optimization problem is how to allocate the connected BS and the available sub-channel to the user, and the optimization objective is to meet the maximization of the system sum rate. Therefore, when the variable ${\Theta}$ is fixed, problem (16) can be written as

\begin{equation}
\begin{split}
&\max_{(S,X)}{\rm{R(S,X)}} = \sum\limits_W {\sum\limits_{{c_i} \in C} {{R^{{c_i}}}} }\\
&\rm {s.t.} \\
 (a)&{X_{{c_i}}^{{n_i}}} \in \{ 0,1\} ,  \forall {n_i},{c_i}, \\
 (b)&\sum\limits_{{n_i} \in N} {X_{{c_i}}^{{n_i}}}  \in \left\{ {0,1} \right\}   ,\forall {n_i},{c_i},\\
 (c)&{X_{{c_i}}^{{c_j}}} \in \left\{ {0,1} \right\},\forall {n_i},{c_i},\\
(d)&{B_{{n_i}}^j} \ge 0, \forall {n_i},{c_i}.
\end{split}
\label{eq17}
\end{equation}

\emph{2) RIS Discrete Phase Shift Optimization:} under the condition that the connected BS and available sub-channel allocated by the user, namely ${(S,X)}$ is fixed, the most appropriate phase shift is selected for each RIS element among ${{2^e}}$ variable provided by quantization bit of ${e}$, so as to maximize the system sum rate. Therefore, when the variables ${(S,X)}$ are fixed, problem (16) can be written as

\begin{equation}
\begin{split}
&\max_{\Theta}R(\Theta ) = \sum\limits_W {\sum\limits_{{c_i} \in C} {{R^{{c_i}}}} }\\
&\rm {s.t.} \\
(a)&{q_{{l_x},{l_z}}} = {e^{j{\theta _{{l_x},{l_z}}}}},{\theta _{{l_x},{l_z}}} = \frac{{2\pi {m_{{l_x},{l_z}}}}}{{{2^e} - 1}},\\
&   {m_{{l_x},{l_z}}} = \{ 0,1,...,{2^e} - 1\} ,1 \le {l_x},{l_z} \le N.
\end{split}
\label{eq18}
\end{equation}

\subsection{Coalition Game Algorithm}\label{S4-2}
In our model, each user will select a network and access it, and users accessing the same BS will form a coalition. In our research, assuming there are ${N}$ BSs and ${P}$ users, the ${P}$ users will be assigned to form ${N}$ coalitions. These coalitions can be expressed as ${C = \{ {c_1},{c_2},...,{c_N}\} }$, where the intersection of any two coalitions is empty, i.e, ${{c_x} \cap {c_{x'}} = \emptyset }$, where ${x,x' \in N}$. In our HCN model, for user p, the user can access ${{n_p}}$ BSs.

In the HCN network, the number of users that initially access one BS ${{c_N}}$ is ${\left| {{c_N}} \right|}$, ${{c_N} \in C}$. According to (14), the sum uplink throughput of the BS ${{c_N}}$, that is utility function of the coalition, is referred to as:

\begin{equation}\label{eq19}
R({c_N}) = \sum\limits_{i = 1}^{\left| C \right|} {{R^{{c_i}}}} . 
\end{equation}

Assuming that a user has ${{n_p}}$ selectable coalitions, the potential coalitions of user ${p}$ can be expressed as ${{C_p} = \{ c_p^1,c_p^2,...,c_p^{{n_p} - 1}\} }$ except for coalition ${{c_N}}$ that the user belongs. According to (19), the utility function of the user  ${p}$'s potential coalitions can be expressed as

\begin{equation}\label{eq20}
R({C_p}) = \{ R(c_p^1),R(c_p^2),...,R(c_p^{{n_p} - 1})\} .
\end{equation}

Because the sub-channel number of each BS is limited, the co-channel interference will occur when the users' number in the coalition exceeds the sub-channel number of BS. The more users, the more severer the interference. In addition, the co-channel interference across BSs will also be generated among BSs on the same frequency.

According to the above analysis, when there are many users in a BS, due to severe interference, the utility function ${R({c_N})}$ of the coalition will not have excellent performance. Since user ${p}$ has ${{n_p} - 1}$ potential coalitions to choose, for the coalition ${{c_N}}$, it is difficult to form a stable coalition. User ${p}$ may leave and choose to join other coalitions, resulting in certain coalitions may be empty and there being no user. In this section, each system model user will construct the coalition as a game player. The joining or leaving of each user is important to the utility function ${R({c_N})}$ of a coalition.

Von Neumann \emph{et al.} were the first to propose and verify the concept of transferable utility in the coalition game \cite{Neumann}. In the model of this paper, the transferable utility coalition of user ${p}$ is denoted as ${({c_N},p,{C_p})}$, that ${{c_N}}$ is the coalition that user ${p}$ is in, and ${{C_p}}$ is a potential coalition group that user ${p}$ can choose to join. For user ${p}$, it is a strategy to choose which coalition to join based on the total utility of the system. The strategy will depend on the coalition utility performance of ${{c_N}}$ and ${{C_p}}$.

The goal of the coalition game algorithm is to obtain the network selection strategy of each user, that is the resource allocation scheme. We can find that user ${p}$ has ${{n_p} - 1}$ potential coalitions to choose. Therefore, to evaluate the coalitions that user ${p}$ chooses to join, we next introduce the concept of the preference relationship \cite{prefer}.

For user ${p}$, the preference relationship means which coalition that user ${p}$ prefers, and it is defined as ${{ \Rightarrow _p}}$. In coalitions constructed by the HCN system, user ${p}$ can choose to stay in the coalition ${{c_N}}$, or switch to the potential coalition ${{C_p}}$ according to the preference relationship. For any user ${p}$, ${{c_N}{ \Rightarrow _p}c_p^x}$ means that for being a member of the coalition ${{c_N}}$, user ${p}$ would prefer to be a member of the potential coalition ${c_p^x}$, where ${c_p^x \in {C_p}}$. For user ${p}$, the preference relation for selecting coalitions is proposed \cite{chenyali2,switch}, and is expressed as
\begin{equation}\label{eq21}
{c_N}{ \Rightarrow _p}c_p^x \Leftrightarrow R({c_N}\backslash p) + R(c_p^x) > R({c_N}) + R(c_p^x\backslash p),
\end{equation}
where ${R({c_N}\backslash p)}$ is the system utility when user ${p}$ is not in the coalition ${{c_N}}$, and ${R(c_p^x)}$ is the system utility when user ${p}$ is in the coalition ${c_p^x}$, and ${R({c_N})}$ is the system utility when user ${p}$ is in the coalition ${{c_N}}$, and ${R(c_p^x\backslash p)}$ is the system utility when user ${p}$ is not in the coalition ${c_p^x}$. The switching can obtain greater system utility, and the difference in system utility is recorded as
\begin{equation}
R(p,{c_N},c_p^x) = (R({c_N}\backslash p) + R(c_p^x)) - (R({c_N}) + R(c_p^x\backslash p)).
\label{eq22}
\end{equation}

After obtaining the preference relationship of user ${p}$, the switching only involves the change of user ${p}$, the coalition ${{c_N}}$, and the coalition ${c_p^x}$. Other users and other coalitions are unchanged. We can denote the updated coalition combination ${C'}$ at this time as
\begin{equation}\label{eq23}
C' = (C\backslash \{ {c_N},c_p^x\} ) \cup \{ {c_N}\backslash p\}  \cup \{ c_p^x \cup \{ p\} \}. 
\end{equation}

The joint optimization process of resource allocation and user association of the HCN in this paper will be described next. First, the initialization operation is performed for ${P}$ users. For any user ${p}$, ${p \in P}$, user ${p}$ will join a coalition evenly and randomly. Until each user belongs to a certain coalition, it means that the initialization operation is completed. Next, according to the order preset by the user set ${P}$, each user will be performed the switching operation of (22). The coalition ${{c_N}}$ where user ${p}$ is in, and the potential coalition ${c_p^x}$, will be performed the calculation of (22). At this time, user ${p}$ will decide to switch to which potential coalition ${c_p^x}$ or continue to stay in the coalition ${{c_N}}$. After user ${p}$ completes the switching operation, the coalition combination ${C}$ of the HCN at this time is updated, and the switching determination of the next user will be performed.

After a round of switching determination of all users is completed, the HCN system utility value is calculated. The next round of the coalition game optimization will be performed on the result of the round. Suppose the result of one round is the best performance in the last 5 rounds. In that case, the algorithm will stop iterating, jump out of the loop, and finally obtain a resource allocation scheme with local optimality. Algorithm \ref{alg1} summarizes the steps of the coalition game algorithm.

\begin{algorithm}[t]
	\caption{The Coalition Game Algorithm for the Joint Optimization of Resource Allocation and User Association}
	\label{alg1}
	\begin{algorithmic}[1]
		\STATE Given the initial partition coalition result ${{C_{ini}}}$ of ${N}$ BSs and ${P}$ users in HCN evenly and randomly;
		\REPEAT
		\FOR{$p$ in $P$}
		\STATE For user p, make a switching calculation between ${{c_N}}$ and the potential coalition ${c_p^x}$, ${c_p^x \in {C_p}}$;
		\IF {The switching calculation between ${{c_N}}$ and ${c_p^x}$ satisfies ${{c_N}{ \Rightarrow _p}c_p^x}$, and ${R(p,{c_N},c_p^x)}$ is the largest in the ${C_p}$}
		\STATE The user $p$ leaves the current coalition ${{c_N}}$ and joins the new coalition ${c_p^x}$;
		\STATE Update the current coalition combination $C' = (C\backslash \{ {c_N},c_p^x\} ) \cup \{ {c_N}\backslash p\}  \cup \{ c_p^x \cup \{ p\} \} $;
		\ENDIF
		\ENDFOR 
		\STATE Calculate system utility value;
		\UNTIL The system utility value is the best performing system in the last 5 rounds.
	\end{algorithmic}
\end{algorithm}

\subsection{Local Discrete Phase Search Algorithm}\label{S4-3}
When the joint optimization of the user association and the resource allocation of the coalition game algorithm is completed, the optimization problem is described in problem (18). It is necessary to select the most appropriate phase shift for each element in the RIS equipped with each mmWave BS. The optional phase shift range for each element is ${[0,2\pi ]}$, because considering realizability, we choose ${{2^e}}$ discrete variables as the optional phase shifts. We use the local search algorithm shown in Algorithm 2 to solve the optimization problem of ${\Theta}$. When we optimize the RIS equipped with one mmWave BS, the phase shift of RIS elements in the remaining mmWave BS is unchanged. For each element ${\{ {l_x},{l_z}\} }$ in the RIS, we first fixed the phase shift values of the other ${{N^2} - 1}$ elements, then traversed ${{2^e}}$ phase shift values to find the phase shift value that maximizes the sum rate of the mmWave BS system at this time. Then we used the phase shift value as the new phase shift value of element ${\{ {l_x},{l_z}\} }$ for the optimization of other elements. After the optimization of all components of the RIS is completed, the next local search algorithm is carried out on the basis of the optimization results of the previous round, until the absolute value of the difference between the system sum rate and that of the last optimization is less than the threshold ${\varepsilon }$. At this time, the RIS optimization is finished.

\begin{algorithm}[t]
	\caption{The Local Discrete Phase Search Algorithm}
	\label{alg2}
	\begin{algorithmic}[1]
		\REQUIRE
		RIS element number of rows and columns ${N}$, the number of quantization bits ${e}$, the threshold value $\varepsilon  = {e^{ - 3}}$,  the number of rounds optimized ${\gamma }$
		\ENSURE
		${\Theta}^{\ast}$
		\REPEAT
		\FOR {$l_{x}=1:N$}
		\FOR {$l_{z}=1:N$}
		\STATE With the remaining elements fixed, assign ${{2^e}}$ phase shift value to ${{\theta _{{l_x},{l_z}}}}$ and find the phase shift value that maximizes the BS system sum rate and assign it to ${{\theta _{{l_x},{l_z}}}^*}$.
		\ENDFOR
		\ENDFOR
		\STATE $\gamma =\gamma +1$, and back to line 1;
		\UNTIL $\left| {{R^{\gamma  + 1}} - {R^\gamma }} \right| < \varepsilon $
	\end{algorithmic}
\end{algorithm}

\subsection{Proposed Algorithm}\label{S4-4}
Block coordinate descent (BCD) is a more generalization of coordinate descent, which decomposes the original problem into multiple sub-problems by simultaneously optimizing a subset of variables. The order of updates during the descent can be deterministic or random \cite{BCD}. The solution idea of BCD is to optimize the solution for only one variable in each iteration, keeping the remaining variables constant, and then solving alternately.

\begin{algorithm}[t]
	\caption{Block coordinate descent}
	\label{alg3}
	\begin{algorithmic}[1]
		\REQUIRE
		choose ${(x_1^0,...,x_s^0)}$
		\FOR{${k}$=1,2,...}
		
		\FOR{${i}$=1,2,...,s}
		\STATE update ${x_i^k}$ with all other blocks fixed
		\ENDFOR 
		\IF {stopping criterion is satisified}
		\STATE return ${(x_1^k,...,x_s^k)}$.
		\ENDIF
		\ENDFOR 
	\end{algorithmic}
\end{algorithm}
Consider an optimization task as follows
\begin{equation}\label{eq24}
\min \;F({x_1},...,{x_s}) \equiv f({x_1},...,{x_s}) + \sum\nolimits_{i = 1}^s {{r_i}({x_i})} .
\end{equation}

A generic framework for BCD is shown in Algorithm 3. In the general framework of BCD, the most commonly used update scheme is block minimization, i.e., ${x_i^k = \mathop {\arg \min }\limits_{{x_i}} F\left( {x_{ < i}^k,{x_i},x_{ > i}^{k - 1}} \right)}$. For (24), we can use coordinate descent to seek a minimum value, and we start with an initial ${{x^{(0)}}}$ that loops over ${k}$, as described in the following unfolding.

\begin{equation}\label{eq25}
\begin{aligned}
x_1^{(k)} &= \mathop {\arg \min }\limits_{{x_1}} F({x_1},x_2^{(k - 1)},x_3^{(k - 1)},...,x_n^{(k - 1)}),\\
x_2^{(k)} &= \mathop {\arg \min }\limits_{{x_2}} F(x_1^{(k)},{x_2},x_3^{(k - 1)},...,x_n^{(k - 1)}),\\
&...\\
x_n^{(k)} &= \mathop {\arg \min }\limits_{{x_n}} F(x_1^{(k)},x_2^{(k)},x_3^{(k)},...,{x_n}).
\end{aligned}
\end{equation}

\begin{algorithm}[ht]
	\caption{RIS-assisted Coalition Game Optimization Algorithm By BCD}
	\label{alg3}
	\begin{algorithmic}[1]
		\REQUIRE
		$\xi  = 3{e^{ - 3}}$, $\varrho=0$, Randomize $(S,X)$ and ${\Theta}$, the number of quantization bits $e$, RIS element number of rows and columns ${N}$
		
		\STATE update ${(S,X)^{\rho  + 1}}$ with fixed ${\Theta  ^{\rho}}$ using Algorithm \ref{alg1};
		\STATE update ${\Theta ^{\rho  + 1}}$ with fixed ${(S,X) ^{\rho}}$ using Algorithm \ref{alg2};
		\IF {$\left| {{R^{\rho  + 1}} - {R^\rho }} \right|/{R^\rho } < \xi$}
		\STATE $R^{\ast}=R^{(\varrho+1)}, \Theta ^{\ast}= \Theta ^{(\varrho+1)}, (S,X)^{\ast}= (S,X)^{(\varrho+1)}$;
		\STATE \textbf{Output} $R^{\ast},{\Theta}^{\ast},(S,X)^{\ast}$;
		\ELSE
		\STATE $\varrho=\varrho+1$, and back to line 1;
		\ENDIF
	\end{algorithmic}
\end{algorithm}
Next, we will discuss the BCD applied in Algorithm 4. 
We first perform initialization at the moment ${\rho  = 0}$, including user resource allocation ${(S,X)}$ and RIS phase ${\Theta }$.
In the next step, the two sub-problems of the section \ref{S4-1} decomposition are computed using BCD, by Algorithm 1 and Algorithm 2. The results are compared with the values calculated in the previous calculation, as shown in step 3 of Algorithm 4.
The first sub-problem is to fix phase ${\Theta }$ of RIS and optimize the user resource allocation scheme ${(S,X)}$ to obtain the local optimal solution; the second is to fix the user resource allocation scheme ${(S,X)}$ and optimize phase ${\Theta }$ of RIS. 
Finally, we determine whether the end-of-iteration condition is satisfied. If the iteration does not end, the next alternate iteration of optimization is performed using BCD.


\subsection{Convergence Analysis}\label{S4-5}

A note on the convergence of Algorithm 1 \cite{Convergence}. Theorem 1: Starting from any initial coalition structure ${{C_{ini}}}$, after a series of exchange operations, the coalition formation algorithm of Algorithm 1 will converge to a final network partition ${{C_{fin}}}$, which is formed by a series of disjoint coalitions.
Proof: By examining the priorities defined in (21), it can be seen that in Algorithm 1, each exchange operation will make user ${p}$ visit an unvisited partition or remain in the present partition by adopting a new policy. As a result, some partitions may be updated to a smaller set of users or even disappear. On the one hand, since there are only ${N}$ BSs, the system will form at most ${N}$ partitions; on the other hand, the number of partitions for a given set ${C}$ of users is the Bell number \cite{switch}. Therefore the sequence of exchange operations will definitely be terminated and converge to the final partition ${{C_{fin}}}$. Therefore, the coalition formation algorithm of Algorithm 1 is convertible.

A note on the convergence of Algorithm 2. The number of discrete phase shifts in Algorithm 2 is finite, allowing the maximization sum rate problems to be bounded and ensuring that the output results converge. Algorithm 2 aims to maximize the system sum rate and obtain ${{\Theta ^{ * (\rho  + 1)}}}$ corresponding to ${{(S,X)^ * }}$. ${{(S,X)^ * }}$ is the result of the user resource allocation obtained by Algorithm 1. The optimal phase shift ${{\Theta ^{ * (\rho  + 1)}}}$ must be greater than or equal to the initial value of ${\Theta }$, which is the initial value of ${{\Theta ^*}}$ obtained by Algorithm 4 after the ${{\rho _{th}}}$ iteration. Finally Algorithm 4 satisfies ${R({(S,X)^{ * (\rho  + 1)}},{\Theta ^{ * (\rho  + 1)}}) \ge R({(S,X)^{ * (\rho)}},{\Theta ^{ * (\rho )}})}$. Therefore, Algorithm 4 is also convergent.

\section{Performance Evaluation}\label{S5}

In this section, with the configuration of various system parameters, the RIS-assisted coalition game optimization algorithm by BCD in Section \ref{S4} will be used to experiment with the resource allocation scheme in the HCN wireless networks assisted by RIS to verify the performance of the proposed algorithm. The performance of the proposed algorithm is compared with several algorithms, and the obtained simulation results are analyzed.

\subsection{Simulation Setup}\label{S5-1}
\begin{table*}[h!t]
	
	\normalsize
	\centering
	\captionsetup{font={normalsize},labelsep=newline, textfont=sc}
	\caption{hcn simulation parameter configuration}
	\label{tab2}
\begin{tabular}{@{}cccccc@{}}
	\toprule
	Parameter                             & Symbol & ${mmWave}$                 & ${5G(1)}$      & ${5G(2)}$      & ${4G}$         \\ \midrule
	Frequency                   & ${F}$     & 26/27/28/29 GHz        & 2.5 GHz       & 4.8 GHz       & 1.9 GHz        \\
	Sub-channel   bandwidth                & ${W}$     & 14.4 MHz                & 3.6 MHz     & 7.2 MHz     & 1.8 MHz     \\
	Transmission   power                  & ${P_c}$     & 21 dBm                  & 26 dBm      & 26 dBm      & 23 dBm      \\
	Noise   power density                 & ${N_0}$      & -174 dBm/Hz             & -174 dBm/Hz & -174 dBm/Hz & -174 dBm/Hz \\
	Path loss   exponent                  & ${n}$      & 2.1                    & 3.8        & 3          & 3.8        \\
	Antenna   gain of UE                 & ${G_u}$      & Depends on the ${\theta }$ & 3 dBi       & 3dBi       & 0.5 dBi     \\
	Antenna   gain of BS                 & ${G_b}$      & Depends on the ${\theta }$ & 25 dBi      & 25dBi      & 13 dBi      \\
	covering   radius                     & ${R}$      & 150 m                    & 350 m        & 300 m        & 1500 m      \\
	Standard   deviation of shadow fading & ${\sigma }$      & 4                      & 6          & 5          & 8          \\\bottomrule
\end{tabular}
\end{table*}

This section considers several BSs and users in a cell. The 4G BS is in the cell center, and other BSs of different network types are distributed around the 4G BS. Since the parameter configuration of various networks is not the same, users' performance accessing different networks will also vary. Table \ref{tab2} at the top of the next page summarizes the parameter settings for the various network types presented in the simulation.

Unlike the cellular networks, the mmWave networks use a widely used practical directional antenna model, which is a Gaussian main lobe on a linear scale with a constant side lobe level \cite{Niummwave}. The model can be expressed as ${G(\theta )}$ in ${dB}$, denoted as (26). ${\beta  = 0.001}$ that ${\beta}$ means the probability of the mmWave connection interruption, and the MUI factor ${\rho  = 1}$ that ${\rho}$ means the mmWave interference among different links.
\begin{equation}\label{eq26}
G(\theta)=\left\{\begin{array}{ll}
G_{0}-3.01 \cdot\left(\frac{2 \theta}{\theta_{-3 {dB}}}\right)^{2}, & 0^{\circ} \leq \theta \leq \theta_{{ml}} / 2, \\
G_{{sl}}, & \theta_{{ml}} / 2 \leq \theta \leq 180^{\circ},
\end{array}\right.
\end{equation}
where ${\theta }$ represents an arbitrary angle  within the range ${[0^\circ ,180^\circ ]}$. In the mmWave network, it is the angle formed by the three positions of the interfering device, the target device, and the BS on the BS side. ${{\theta _{ - 3dB}}}$ denotes the angle of the half-power beam width, that ${{\theta _{ - 3dB}} = 30^\circ }$ and ${{\theta _{ml}}}$ represents the main lobe width in degrees. The relationship between ${{\theta _{ - 3dB}}}$ and ${{\theta _{ml}}}$ is ${{\theta _{ml}} = 2.6 \cdot {\theta _{ - 3dB}}}$. ${{G_0}}$ is the maximum antenna gain, and ${{G_{sl}}}$ refers to the side lobe gain, which can be denoted, respectively, as

\begin{equation}\label{eq27}
{G_0} = 10\log \left(\frac{{1.6162}}{{\sin ({\theta _{ - 3dB}}/2)}}\right)^2,
\end{equation}

\begin{equation}\label{eq28}
{G_{sl}} =  - 0.4111 \cdot \ln ({\theta _{ - 3dB}}) - 10.579.
\end{equation}

At the same time, we set RIS on the north side of the communication coverage radius of each mmWave BS. The RIS is a ${N \times N}$ uniform planar array programmable elements, where $N=4$ and quantization bits ${e =3}$. ${\hat{h}_{NLoS, l_{z}, l_{y}}^{r_{i}, t_{i}} \sim \mathcal{C N}(0,1)}$ is the small-scale fading. It obeys the Nakagami-mi distribution with parameters, that ${{m_i} = 3}$ is the fading depth parameter and ${{w_i} = 1/3}$ is the average power in the fading signal.

\begin{figure}[t]
	\begin{center}
		\includegraphics[height=2.5in]{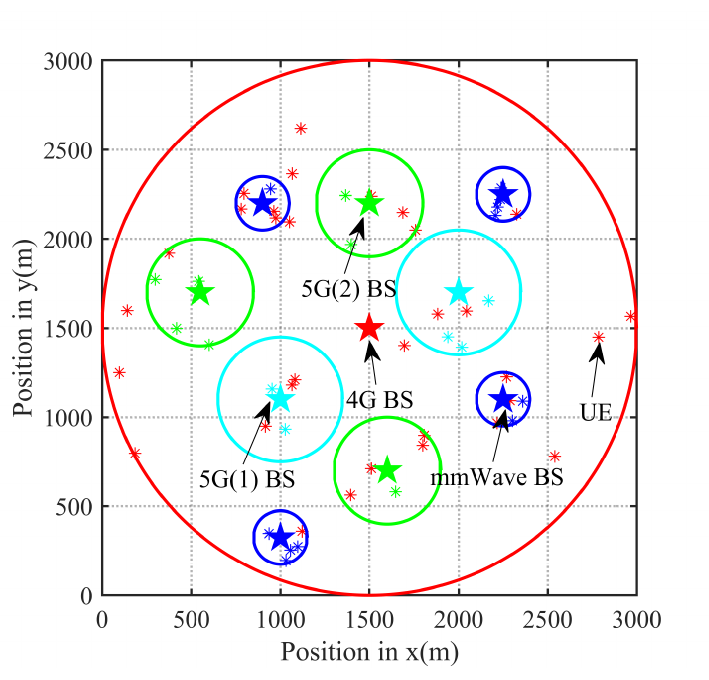}
	\end{center}
	\caption{Initialized HCN model consisting of 10 BSs and 55 users} \label{fig3}
\end{figure}

\begin{figure}[t]
	\begin{center}
		\includegraphics[height=2.5in]{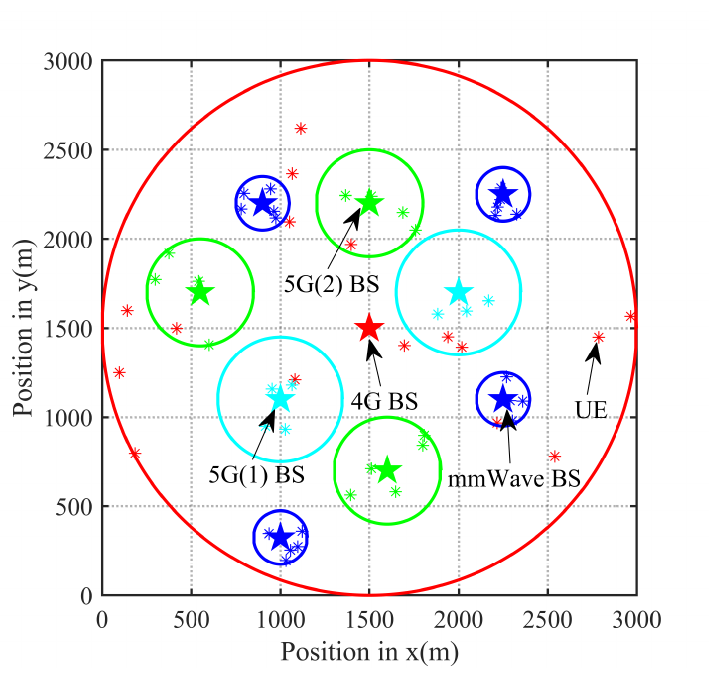}
	\end{center}
	\caption{HCN model optimized by the proposed algorithm consisting of 10 BSs and 55 users} \label{fig4}
\end{figure}

To explain the distribution of BSs and users in the HCN model. Fig. \ref{fig3} shows that after setting the location of BSs, a network model composed of 55 users is generated, and the positions of various BSs and users are marked. Fig. \ref{fig4} shows the scheme of the resource allocation and user association that the proposed algorithm optimizes the model of Fig. \ref{fig3}. Five-pointed stars represent the BSs, red represents the 4G BS, dark blue represents the mmWave BSs with four different frequencies, light blue represents 5G(1) BSs, and green represents 5G(2) BSs, and asterisks represent users. The asterisk's color indicates which BS the user is connected to.
Take the BS represented by the blue pentagram in the upper left corner in Fig. 3 as an example for illustrating the optimization effect. We can see 4 red asterisks in the coverage area of the BS, representing four users who were in the coverage area of both 4G and mmwave BSs before the optimization, and established communication with the 4G BS. In Fig. 4, which is optimized by the proposed algorithm, we can see that these 4 users have changed from the previous red to blue color, indicating that they have established a connection with the mmwave BS. The communication rate has been effectively improved. The above results show that the user resource allocation and user association in HCN are optimized.

To further evaluate the advantage of the proposed algorithm (PA) in Algorithm 4 in terms of the system sum rate, we compare four algorithms with the PA. These four algorithms include the coalition game algorithm (CGA), RIS optimization (RO), random algorithm (RA), and the cellular coalition games algorithm (CCGA):
\begin{itemize}
	\item \textbf{Coalition game Algorithm (CGA)}. The difference between the coalition game algorithm and PA is that RIS is not installed at the mmWave BS, and there is no optimization of the RIS phase shift. Only the coalition game algorithm of Algorithm \ref{alg1} is used to optimize resource allocation and user association.
	\item \textbf{RIS optimization (RO)}. The difference between the RIS optimization algorithm and PA is that the RIS element of the mmWave BS is only optimized by the local discrete phase search algorithm of Algorithm \ref{alg2}, but the resource allocation and user association are not optimized.
	\item \textbf{Random algorithm (RA)}. In the wireless resource scheduling method, the HCN system randomly selects a potential BS with an equal probability of users. Therefore, the selection of each user has no connection with the selection of other users, and the selection is entirely random. The mmWave BS does not have RIS installed. In this paper, the final RA result will be the average of 100 RA results.
	\item \textbf{Cellular Coalition Games Algorithm (CCGA)}. In this strategy, the mmWave BS will not provide network connection services and isn't equipped with RIS. The users in the coverage area of the mmWave BSs can join other potential BSs. After the connection is initialized, the coalition games algorithm in Algorithm 1 will be carried out.
\end{itemize}

\subsection{Complexity Analysis}\label{S5-2}
The complexity of the proposed algorithm RIS-assisted Coalition Game Optimization Algorithm By BCD is related not only to the number of iterations, but also to the complexity of the resource allocation sub-problem and the phase shift optimization sub-problem. The number of iterations is set to ${{N_3}}$ to achieve the convergence condition ${\left| {{R^{\rho  + 1}} - {R^\rho }} \right|/{R^\rho } < \varepsilon }$. For the resource allocation, the number of coalition games until convergence, and the number of users switching under multi-BS coverage in each coalition game iteration introduces complexity. Among ${P}$ users, the number of users involved under multi-BS coverage is set to ${{P_1}}$, and the number of users under single-BS coverage is set to ${{P_2}}$. Let ${{N_1}}$ be the number of iterations of the resource allocation sub-problem to achieve convergence in Algorithm 1. Since only the users in ${{P_1}}$ will be involved in switching operations in each iteration, and there is at most one switching operation, the complexity of the resource allocation sub-problem is ${o({N_1} * {P_1})}$. For each RIS element ${\{ {l_x},{l_z}\} }$ in the phase shift optimization sub-problem, the local search algorithm keeps the phase shifts of the remaining elements constant, selects the optimal one from the ${{2^e}}$ phase shift values, and updates the value of ${{\theta _{{l_x},{l_z}}}}$. Since RIS is a ${N \times N}$ planar array, the complexity of RIS in each iteration is ${o({N^2} * {2^e})}$. 
Because there are multi-BS in the model of this paper, among which all mmwave BSs are equipped with RISs, with ${M}$ mmwave BSs, and the number of phase shift optimization iterations of RISs is ${N_2^k}$ to satisfy the convergence condition ${\left| {{R^{\gamma  + 1}} - {R^\gamma }} \right| < \varepsilon }$ in Algorithm 2. Thus the complexity of the phase shift optimization sub-problem is ${o\left( {\sum\limits_{k = 1}^M {N_2^k*\left( {{N^2}*{2^e}} \right)} } \right)}$. The final complexity of the proposed algorithm in this paper is ${o\left( {{N_1}*{P_1} + \sum\limits_{k = 1}^M {N_2^k * \left( {{N^2} * {2^e}} \right)} } \right)}$. The complexity analysis of the CGA, RO, and CCGA algorithms is similar to that of the PA algorithm for ${o\left( {{N_1}*{P_1}} \right)}$, ${o\left( {\sum\limits_{k = 1}^M {N_2^k * \left( {{N^2} * {2^e}} \right)} } \right)}$, and ${o\left( {{N_1}*{P_1}} \right)}$, respectively.


\subsection{Performance Comparison}\label{S5-3}
We evaluate the proposed algorithm's performance by changing the system parameters' configuration. Next, we compare the results of the proposed algorithm with the simulation of the other four algorithms.

Fig. 5 shows the simulation result by changing the users' number in each BS' coverage area, with the condition that the sub-channel number provided by BSs is fixed. The sub-channel number of the 4G BS is 6, and that of other BSs is 4. RIS is set up with ${N = 4}$, ${e = 3}$. The users' number in each BS is shown in Table \ref{tab3}, that GU represents the order of the user group. It can be seen that the users' number in each BS in Table \ref{tab3} increases gradually as the value of GU increases. When GU = 1, the system sum rate of PA is about 47\% higher than CGA's. This is because the PA algorithm considers the phase optimization of RIS based on CGA, which effectively improves the communication quality of mmWave and reduces interference. When GU = 1, the system sum rate of PA is about 58\% higher than RO's. This is because the PA algorithm takes into account the coalition game algorithm, which is jointly optimized for resource allocation and user association based on the RO algorithm. It reasonably allocates sub-channels provided by BS and access to each user, which improves the overall system sum rate. However, CCGA, without mmWave network service, and RA, without mmWave and RIS, have a significantly lower system sum rate than the other three algorithms. In the later stage, when the users' number increases to a certain value, the co-channel interference will become more and more severe due to the insufficient sub-channels provided by the BS, so the system sum rate will tend to be stable.

\begin{table}
	\centering
	\captionsetup{font={normalsize},labelsep=newline, textfont=sc}
	\caption{combination of user/sub-channel in bs }
	\label{tab3}
	\begin{tabular}{@{}cccccccc@{}}
		\toprule
		\multirow{2}{*}{\diagbox{BS}{GU/GC}} & \multirow{2}{*}{1} & \multirow{2}{*}{2} & \multirow{2}{*}{3} & \multirow{2}{*}{4} & \multirow{2}{*}{5} & \multirow{2}{*}{6} & \multirow{2}{*}{7} \\
		&   &   &   &   &   &   &    \\ \midrule
		${4G}$        & 4 & 5 & 6 & 7 & 8 & 9 & 10 \\
		${other}$  & 2 & 3 & 4 & 5 & 6 & 7 & 8  \\\bottomrule
	\end{tabular}
\end{table}

\begin{figure}[t]
	\begin{center}
		\includegraphics[height=2.25in]{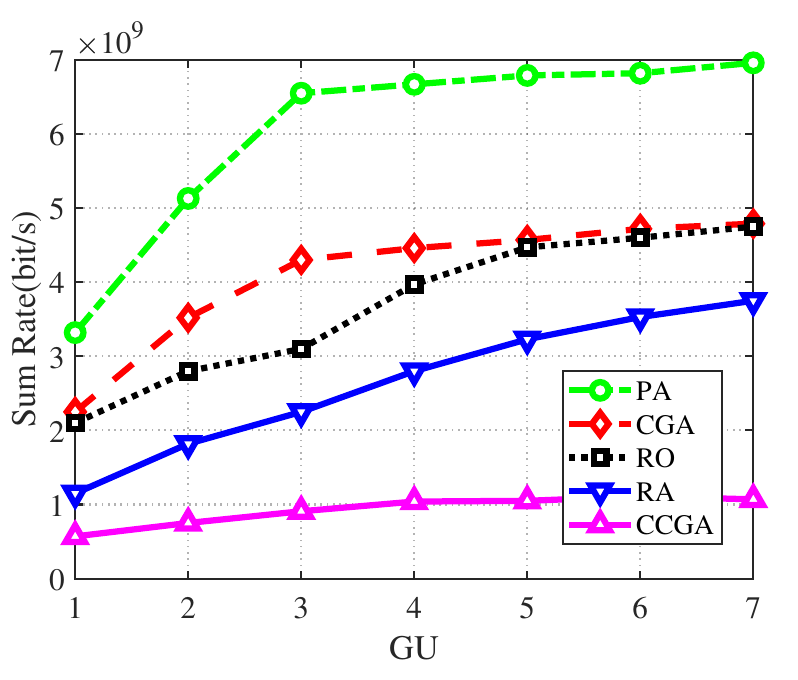}
	\end{center}
	\caption{System sum rate comparison of five resource allocation algorithms with different user group} \label{fig5}
\end{figure}

\begin{figure}[t]
	\begin{center}
		\includegraphics[height=2.25in]{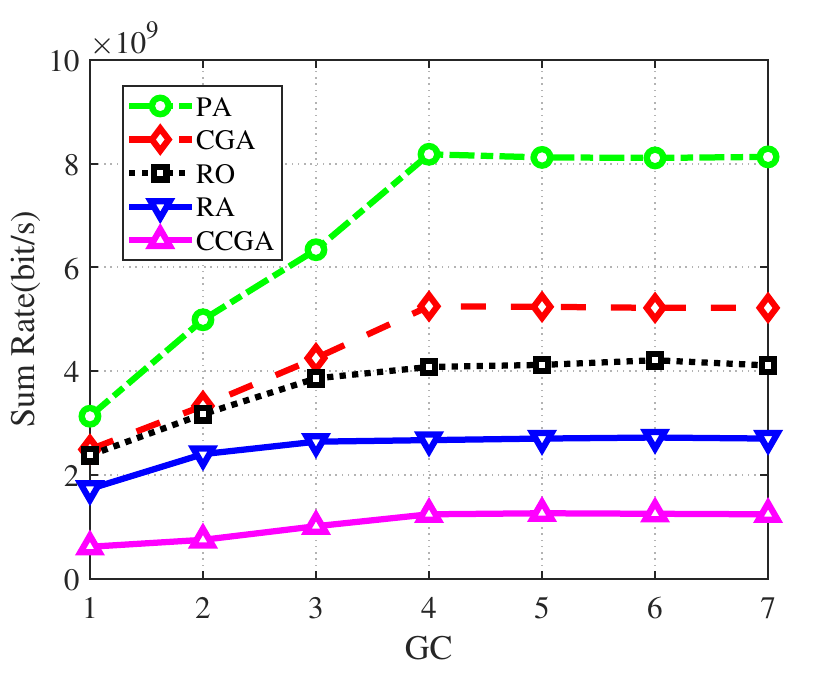}
	\end{center}
	\caption{System sum rate comparison of five resource allocation algorithms with different sub-channel group} \label{fig6}
\end{figure}

Fig. \ref{fig6} shows the simulation result by changing the sub-channel number provided by BSs, in the case of fixing the users' number in the coverage area of each BS. The user number of the 4G BS is 7, and that of other BSs is 5. RIS is set up with ${N = 4}$, ${e = 3}$. It can be seen that the sub-channel number of each BS in Table \ref{tab3} increases gradually as the value of GC increases, that GC represents the order of the sub-channel group of each BS. 
When GC = 1, the system sum rate of PA is about 26\% higher than that of CGA and 32\% higher than that of RO, because PA is an alternate iteration algorithm of CGA and RO by BCD, taking into account both resource allocation and RIS phase-shift optimization. When the number of sub-channels exceeds the users' number, the co-channel interference will not exist, and the system sum rate will not increase.

\begin{figure}[t]
	\begin{center}
		\includegraphics[height=2.25in]{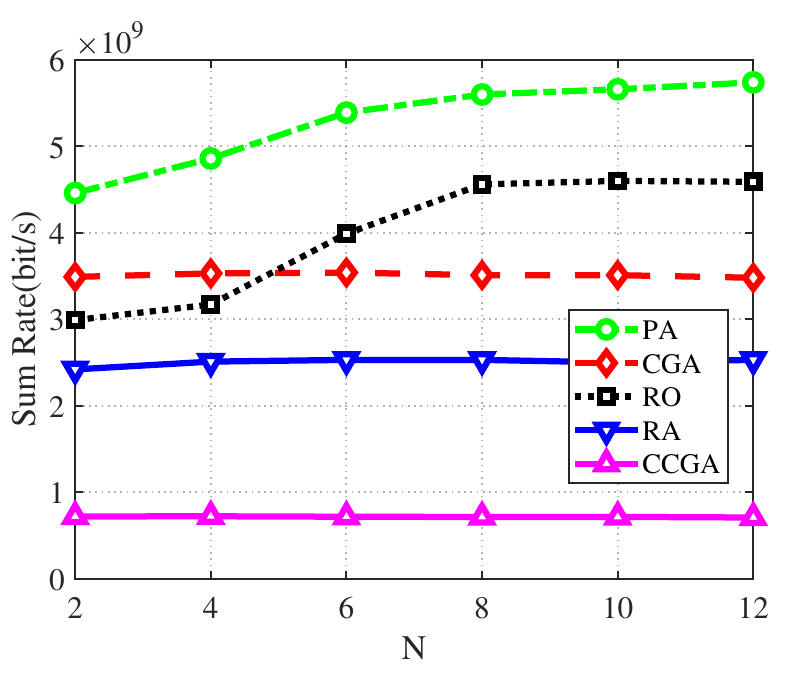}
	\end{center}
	\caption{System sum rate comparison of five resource allocation algorithms with different ${N}$} \label{fig7}
\end{figure}

\begin{figure}[t]
	\begin{center}
		\includegraphics[height=2.25in]{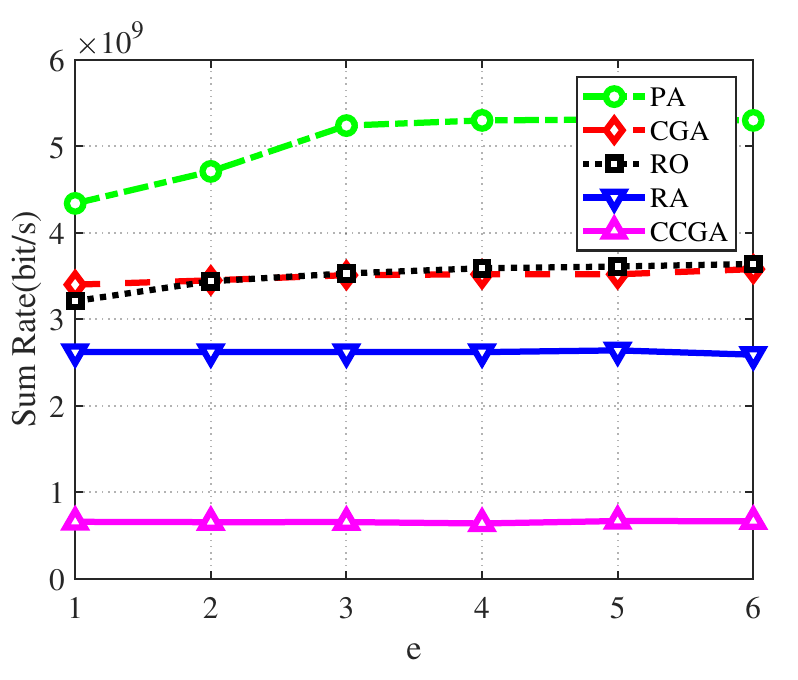}
	\end{center}
	\caption{System sum rate comparison of five resource allocation algorithms with different ${e}$} \label{fig8}
\end{figure}

\begin{figure}[t]
	\begin{center}
		\includegraphics[height=2.25in]{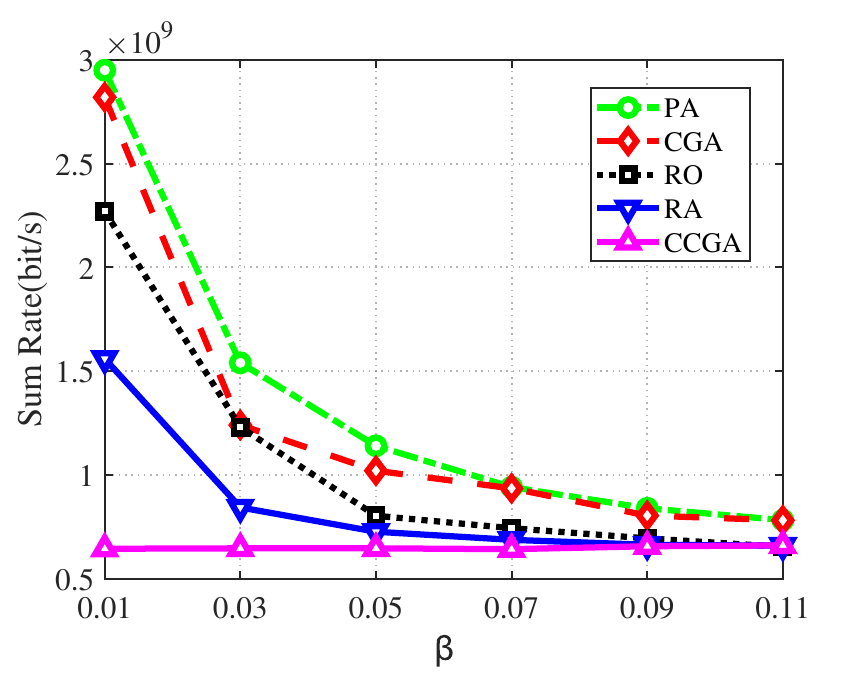}
	\end{center}
	\caption{System sum rate comparison of five resource allocation algorithms with different ${\beta }$} \label{fig9}
\end{figure}

\begin{figure}[t]
	\begin{center}
		\includegraphics[height=2.25in]{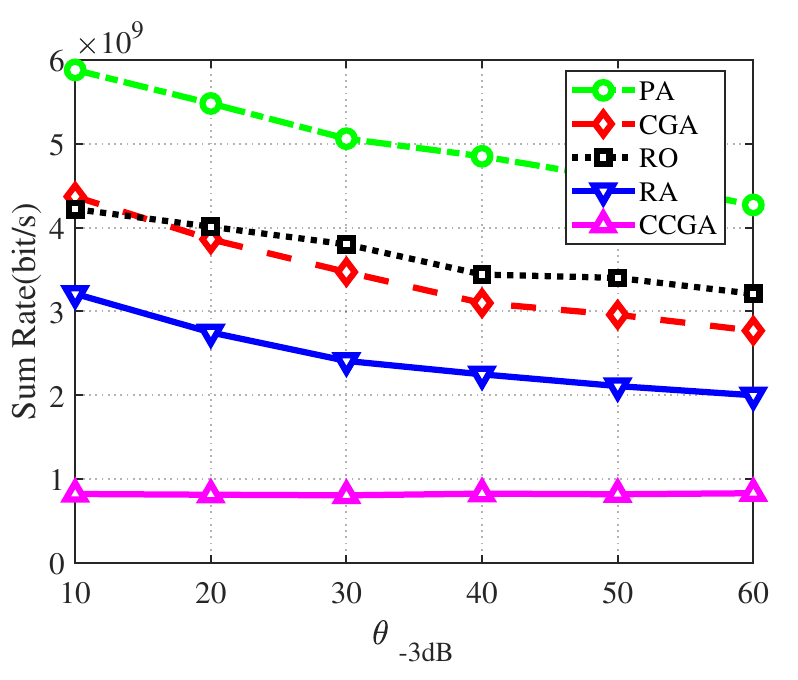}
	\end{center}
	\caption{System sum rate comparison of five resource allocation algorithms with different ${\theta _{-3dB}}$} \label{fig10}
\end{figure}

\begin{figure}[t]
	\begin{center}
		\includegraphics[height=2.25in]{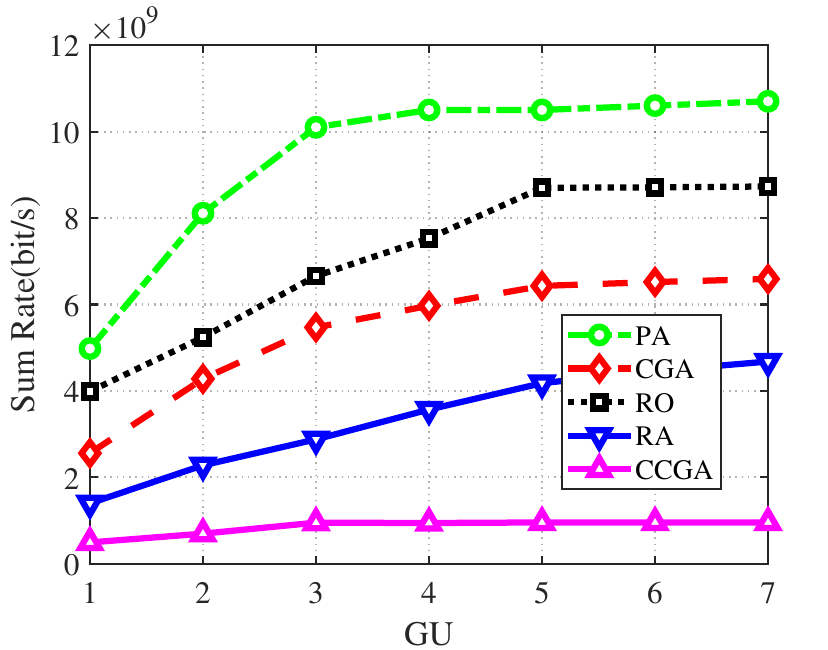}
	\end{center}
	\caption{System sum rate comparison of 6G HCN network with different user groups under five resource allocation algorithms} \label{fig11}
\end{figure}

Fig. 7 shows the simulation results of the system sum rate obtained by changing the number of elements ${N}$ in each row and column of RIS. The user number of the 4G BS is 7 and that of other BSs is 5. The sub-channel number of the 4G BS is 5, and that of other BSs is 3. The number of RIS quantized bits is ${e = 3}$. It can be seen that the system sum rate of PA and RO increase with the increase of ${N}$, because with the increase of ${N}$, RIS can provide more reflected signals and increase the communication quality of mmWave. In the later stage, although ${N}$ is increasing, the reflected signal provided by RIS tends to saturation, so the system sum rate of PA and RO is stable. The remaining three algorithms do not have RIS installed or provide mmWave service, so the system sum rate will not change with the ${N}$ change.

Fig. 8 shows the simulation results of the system sum rate obtained by changing RIS's quantization bit ${e}$. The user number of the 4G BS is 7, and that of other BSs is 5. The sub-channel number of the 4G BS is 5, and that of other BSs is 3. The element of RIS is ${N=4}$. By analyzing Fig. 8, it can be found that the system sum rate of PA and RO increases with the increase of ${e}$. This is because as ${e}$ increases, the number of phase shifts available to the RIS element increases, and phase shifts satisfying a larger system sum rate will be found. However, in the later stage, when RIS elements have found suitable phase shifts, increasing the number of phase shifts can not effectively improve the quality of mmWave communication. So the system sum rate of PA and RO is stable. The remaining three algorithms do not have RIS installed or provide mmWave service, so the system sum rate will not change with the ${e}$ change.

Fig. 9 shows the system sum rate simulation results for changing the interruption parameter ${\beta }$ of the mmWave network. The user number of the 4G BS is 7, and that of other BSs is 5. The sub-channel number of the 4G BS is 5, and that of other BSs is 3. RIS is set up with ${N = 4}$, ${e = 3}$. Changing ${\beta }$ will affect the mmWave networks. Since CCGA does not provide the mmWave network service, the change of ${\beta }$ does not affect CCGA. 
For the remaining four algorithms, the system sum rate decrease with the increase of ${\beta }$ due to the mmWave network provided. In the latter stage, due to users leaving the mmWave network and choosing to access other networks, and the severe communication loss of mmWave, the system sum rate of the five algorithms will slowly approach in the later stage.

Fig. 10 shows the simulation results of the system sum rate obtained by changing the angle of the half-power beamwidth ${\theta_{-3dB}}$ of the mmWave. The user number of the 4G BS is 7, and that of other BSs is 5. The sub-channel number of the 4G BS is 5, and that of other BSs is 3. RIS is set up with ${N = 4}$, ${e = 3}$. ${\theta_{-3dB}}$ is the angle of the half-power beamwidth adopting the widely used realistic directional antenna model. Because the mmWave network directionally transmits signals, the larger the ${\theta_{-3dB}}$, the wider the coverage area of the mmWave antenna, which will cause more severe interference to other mmWave users. Therefore, it can be seen from Fig. 10 that the system sum rate of the four algorithms, except CCGA, decreases with the increase of ${\theta_{-3dB}}$. When ${\theta_{-3dB}=50}$, the downward trend of the system sum rate slows down because the interference is nearing saturation.

We include a network in the terahertz band in this paper to show the performance of the proposed algorithm in the 6G network. We replace a mmwave BS in the model of Fig. 3 with a terahertz BS at 0.34 THz to avoid peak path loss \cite{Thz6G2}, a bandwidth of 10 GHz, a transmitted power of 26 dBm, a path loss exponent ${n = 2}$ \cite{Thz6G3}, and a shadow fading of 10 dB \cite{Thz6G4}. We use the proposed algorithm to simulate the system sum rate with increasing users for an HCN joined to a Thz network. This is shown in Fig. 11. Fig. 11 shows the system sum rate comparison in an HCN containing terahertz, mmwave, 5G mid-frequency, 5G low-frequency, and 4G networks with the number of users, optimized by five resource allocation algorithms. Adding the terahertz network in Fig. 11 greatly improves the system sum rate compared to Fig. 5. The PA algorithm performs the best due to resource allocation and RIS phase shift optimization. The RO algorithm focuses on the RIS phase shift optimization, which becomes more pronounced due to the addition of the terahertz network compared to Fig. 5. Since the effective band number of BS is fixed, the system sum rate comparison will gradually stabilize as the number of users increases, and co-channel interference becomes more severe.

\subsection{Fairness Analysis}\label{S5-4}
Finally, we use Jain's fairness index, commonly used in research, to examine the fairness of the system \cite{jainfair}, whose expression is given in equation (29).

\begin{figure}[h]
	\begin{center}
		\includegraphics[height=2.25in]{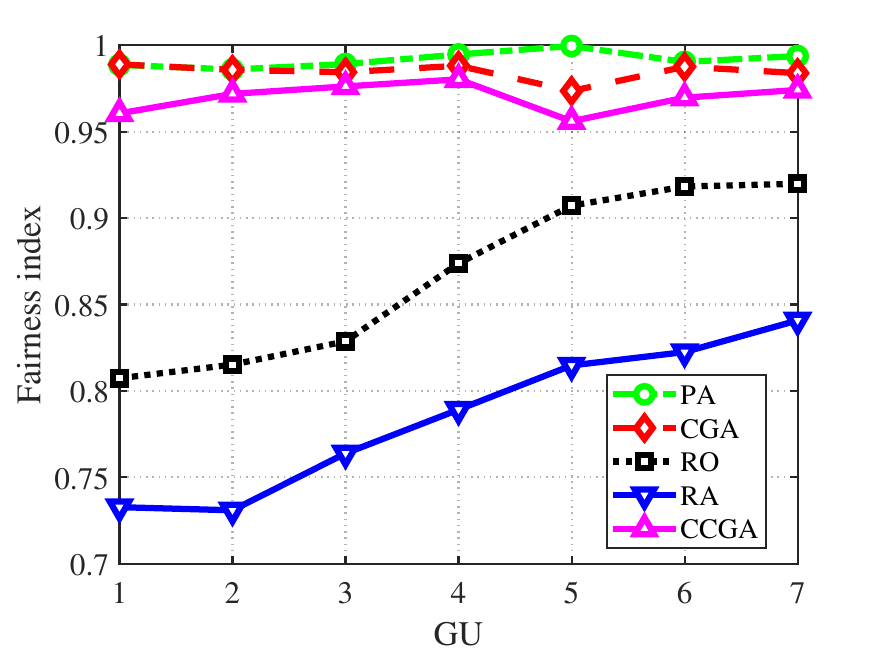}
	\end{center}
	\caption{Fairness performance of five resource allocation algorithms with different user group} \label{fig12}
\end{figure}

\begin{equation}\label{eq29}
FI = \frac{{{{\left( {\sum\nolimits_{i = 1}^n {{R_i}} } \right)}^2}}}{{\left( {n\sum\nolimits_{i = 1}^n {R_i^2} } \right)}},
\end{equation}
where ${{R_i}}$ is the system sum rate of the ${{i_{th}}}$ BS, and there are ${n}$ BSs. The fairness index varies from 0 (worst) to 1 (best).

Analyzing Fig. 12, we can find that the PA, CGA, and CGGA algorithms involving coalition gaming algorithms all obtain higher fairness indices, with PA's being the highest. In contrast, the RO and RA algorithms have slightly worse fairness indexes. It illustrates that the coalition game algorithm can effectively improve the fairness of resource allocation while enhancing system performance. And the PA algorithm exhibits better fairness and obtains higher uplink throughput due to the consideration of RIS phase shift optimization.

\subsection{Compared With the Optimal Solution}\label{S5-5}
\begin{figure}[h]
	\begin{center}
		\includegraphics[height=2.25in]{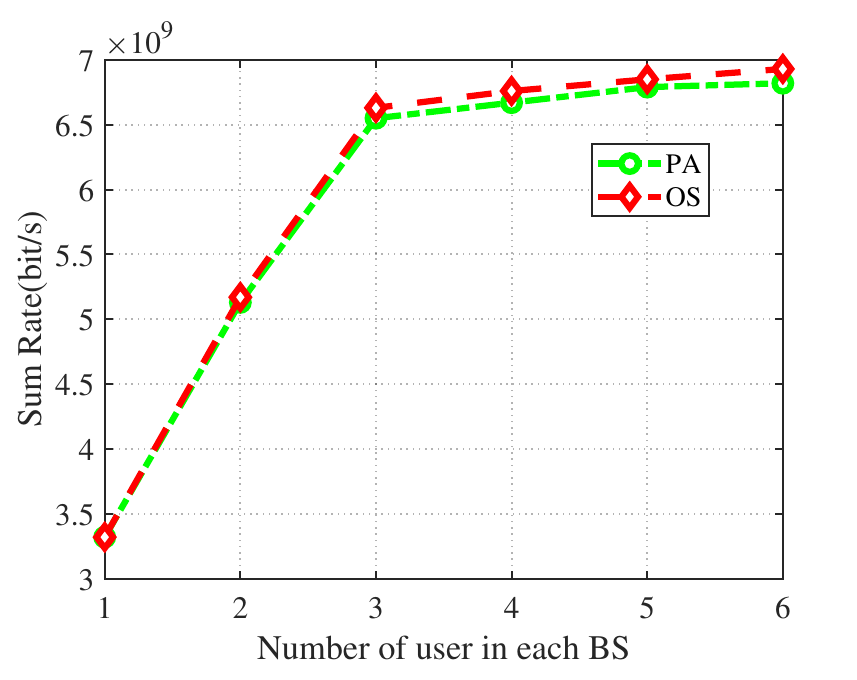}
	\end{center}
	\caption{Comparison of the system sum rate of the proposed algorithm and the traversal algorithm with different users} \label{fig13}
\end{figure}

\begin{figure}[h]
	\begin{center}
		\includegraphics[height=2.25in]{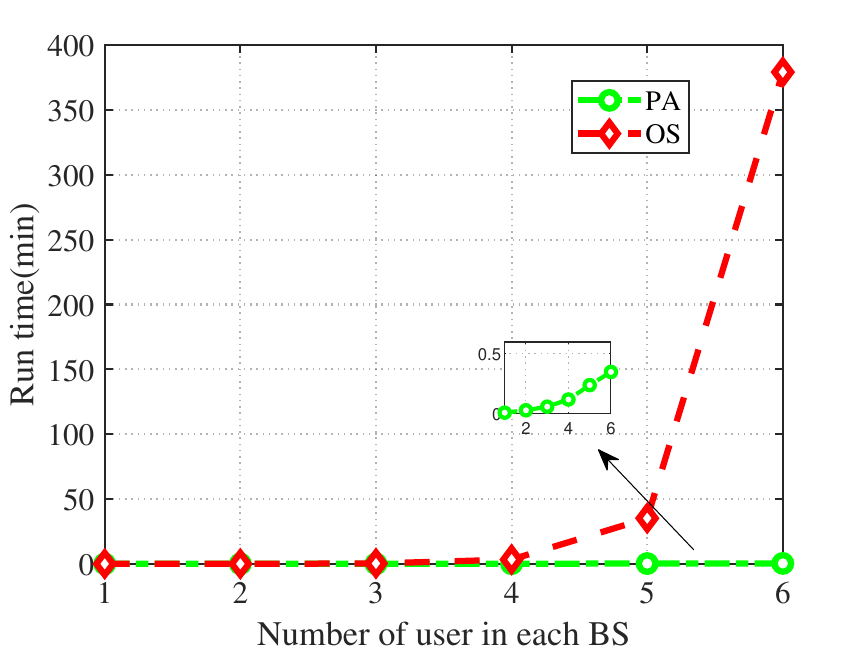}
	\end{center}
	\caption{Comparison of the running time of the proposed algorithm and the traversal algorithm with different users} \label{fig13}
\end{figure}

In this subsection, we compare the performance of the PA algorithm with the performance of the optimal solution (OS) obtained by the traversal method. Given the high complexity of the traversal algorithm, the users of each BS are taken as values from 1 to 6 to obtain the experimental results in Fig. 13 and Fig. 14, where Fig. 13 shows the performance of the PA compared with the traversal algorithm in terms of the system sum rate. From Fig. 13, it can be seen that the system sum rate of PA has an excellent approximation to that achieved by OS. To further demonstrate that the PA converges near the traversal algorithm, we analyzed the simulation results in detail and calculated the average deviation of the results obtained by PA and OS, expressed as follows:

\begin{equation}\label{eq30}
Average{\kern 1pt} {\kern 1pt} {\kern 1pt} {\kern 1pt} {\kern 1pt} {\kern 1pt} {\kern 1pt} Deviation = \frac{1}{6}\sum\limits_{n = 1}^6 {\frac{{{R_{OS}}(n) - {R_{PA}}(n)}}{{{R_{OS}}(n)}}}, 
\end{equation}
where ${ {{R_{OS}}(n)}}$ and ${ {{R_{PA}}(n)}}$ denote the system sum rate obtained by OS and PA, respectively, and ${n}$ is the number of users in each BS. The average deviation between the OS and PA in Fig. 13 is about 0.9\%, which indicates that PA in this paper can achieve the system sum rate close to the optimal solution.

Fig.14 shows the performance comparison of the running time of the PA and the traversal algorithm to obtain the OS, and it can be found that the time required by the PA algorithm is much smaller than that of the traversal algorithm.
Therefore, the analysis of Fig. 13 and Fig. 14 can show that PA achieves a performance close to the optimal solution obtained by the traversal algorithm with low complexity.

\section{Conclusion}\label{S6}
This study investigated user resource allocation in the HCN aided by RIS. We first developed an HCN model that incorporates multiple BSs and frequency bands, including 4G, 5G, mmwave, and terahertz, to align with realistic network scenarios and 6G network concepts. RIS technology was employed to enhance the quality of mmwave signals and mitigate interference.
To address the optimization problem of maximizing the system sum rate, we proposed an algorithm that alternates between optimizing resource allocation and RIS phase shift using the BCD algorithm. This algorithm divides the system optimization problem into two subproblems, which are solved using a coalition game algorithm and a local discrete phase search algorithm, respectively. Unlike previous studies, our proposed algorithm effectively handles the NP-hard nature of the optimization problem and the coupling relationship between the two subproblems.
Simulation results demonstrated that the system performance of mmwave BS significantly improved with the assistance of RIS, leading to a substantial enhancement in the overall system throughput. Compared to other algorithms that lack coalition game optimization, our coalition game algorithm reduced co-channel interference by optimizing resource allocation, thereby improving the system sum rate. As a result, the coalition game algorithm assisted by RIS outperforms other algorithms, achieving near-optimal solution performance with low complexity.
In future work, we intend to focus on exploring the interconnected relationship within HCN and the application of RIS in radio resource management (RRM). We aim to find effective solutions for the coexistence and interoperability of International Mobile Telecommunications (IMT) systems with next-generation networks. In addition, we will further investigate the setup and traffic patterns of 6G HCNs, which include multiple antennas at BS and MU-MIMO, to enhance our study's value and impact.

\small
\bibliographystyle{IEEEtran}
\bibliography{Colation_games_refer}
\bibliographystyle{IEEEtran}

\begin{IEEEbiography}[{\includegraphics[width=1in,height=1.25in,clip,keepaspectratio]{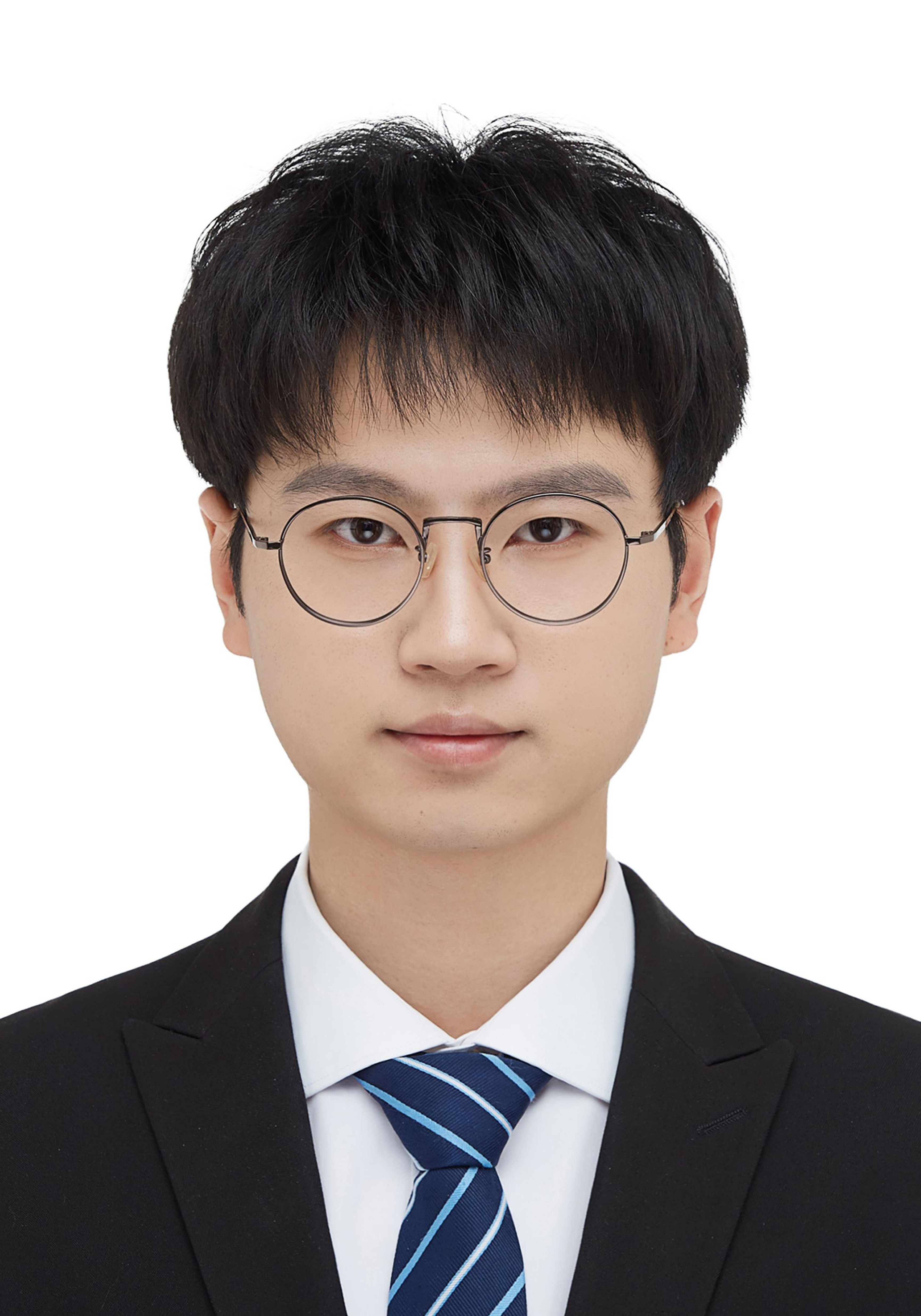}}]{Yuanyuan Qiao}
	 received the B.S. degree in communication engineering from North China Electric Power University, Hebei, China, in 2019 and the M.Eng. degree in electronics and communication engineering from Beijing Jiaotong University, Beijing, China, in 2021. He is currently pursuing the Ph.D. degree with the State Key Laboratory of Rail Traffic Control and Safety, Beijing Jiaotong University, Beijing, China. His current research interests include wireless resource allocation, ultra-reliable low-latency communications, and high-speed railroad communications.
\end{IEEEbiography}

\begin{IEEEbiography}[{\includegraphics[width=1in,height=1.25in,clip,keepaspectratio]{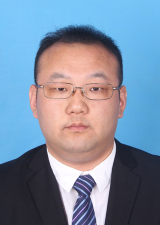}}]{Yong Niu}
	(Senior Member, IEEE) received the B.E. degree in Electrical Engineering from Beijing Jiaotong University, China, in 2011, and the Ph.D. degree in Electronic Engineering from Tsinghua University, Beijing, China, in 2016.
	
	From 2014 to 2015, he was a Visiting Scholar with the University of Florida, Gainesville, FL, USA. He is currently an Associate Professor with the State Key Laboratory of Rail Traffic Control and Safety, Beijing Jiaotong University. His research interests are in the areas of networking and communications, including millimeter wave communications, device-to-device communication, medium access control, and software-defined networks. He received the Ph.D. National Scholarship of China in 2015, the Outstanding Ph.D. Graduates and Outstanding Doctoral Thesis of Tsinghua University in 2016, the Outstanding Ph.D. Graduates of Beijing in 2016, and the Outstanding Doctorate Dissertation Award from the Chinese Institute of Electronics in 2017. He has served as Technical Program Committee member for IWCMC 2017, VTC2018-Spring, IWCMC 2018, INFOCOM 2018, and ICC 2018. He was the Session Chair for IWCMC 2017. He was the recipient of the 2018 International Union of Radio Science Young Scientist Award.
\end{IEEEbiography}

\begin{IEEEbiography}[{\includegraphics[width=1in,height=1.25in,clip,keepaspectratio]{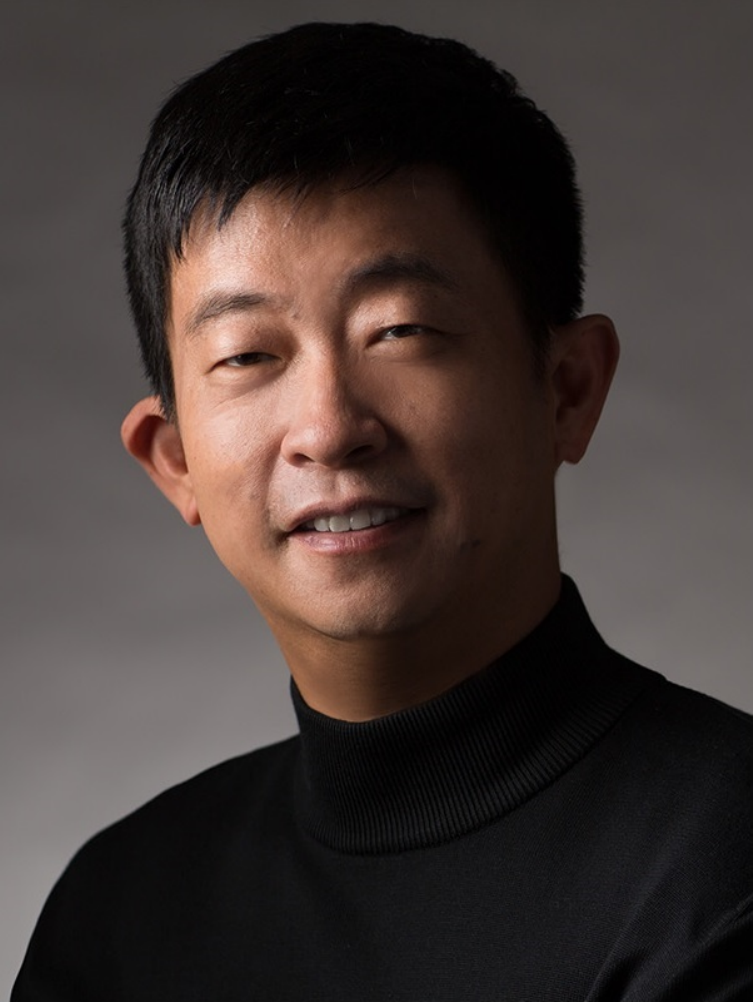}}]{Zhu Han}
(S’01–M’04-SM’09-F’14) received a B.S. degree in electronic engineering from Tsinghua University, in 1997, and M.S. and Ph.D. degrees in electrical and computer engineering from the University of Maryland, College Park, in 1999 and 2003, respectively. 

From 2000 to 2002, he was an R\&D Engineer of JDSU, Germantown, Maryland. From 2003 to 2006, he was a Research Associate at the University of Maryland. From 2006 to 2008, he was an assistant professor at Boise State University, Idaho. Currently, he is a John and Rebecca Moores Professor in the Electrical and Computer Engineering Department as well as in the Computer Science Department at the University of Houston, Texas. Dr. Han’s main research targets on the novel game-theory related concepts critical to enabling efficient and distributive use of wireless networks with limited resources. His other research interests include wireless resource allocation and management, wireless communications and networking, quantum computing, data science, smart grid, security and privacy.  Dr. Han received an NSF Career Award in 2010, the Fred W. Ellersick Prize of the IEEE Communication Society in 2011, the EURASIP Best Paper Award for the Journal on Advances in Signal Processing in 2015, IEEE Leonard G. Abraham Prize in the field of Communications Systems (best paper award in IEEE JSAC) in 2016, and several best paper awards in IEEE conferences. Dr. Han was an IEEE Communications Society Distinguished Lecturer from 2015-2018, AAAS fellow since 2019, and ACM distinguished Member since 2019. Dr. Han is a 1\% highly cited researcher since 2017 according to Web of Science. Dr. Han is also the winner of the 2021 IEEE Kiyo Tomiyasu Award (an IEEE Technical Field Award), for outstanding early to mid-career contributions to technologies holding the promise of innovative applications, with the following citation: ``for contributions to game theory and distributed management of autonomous communication networks."
\end{IEEEbiography}

\begin{IEEEbiography}[{\includegraphics[width=1in,height=1.25in,clip,keepaspectratio]{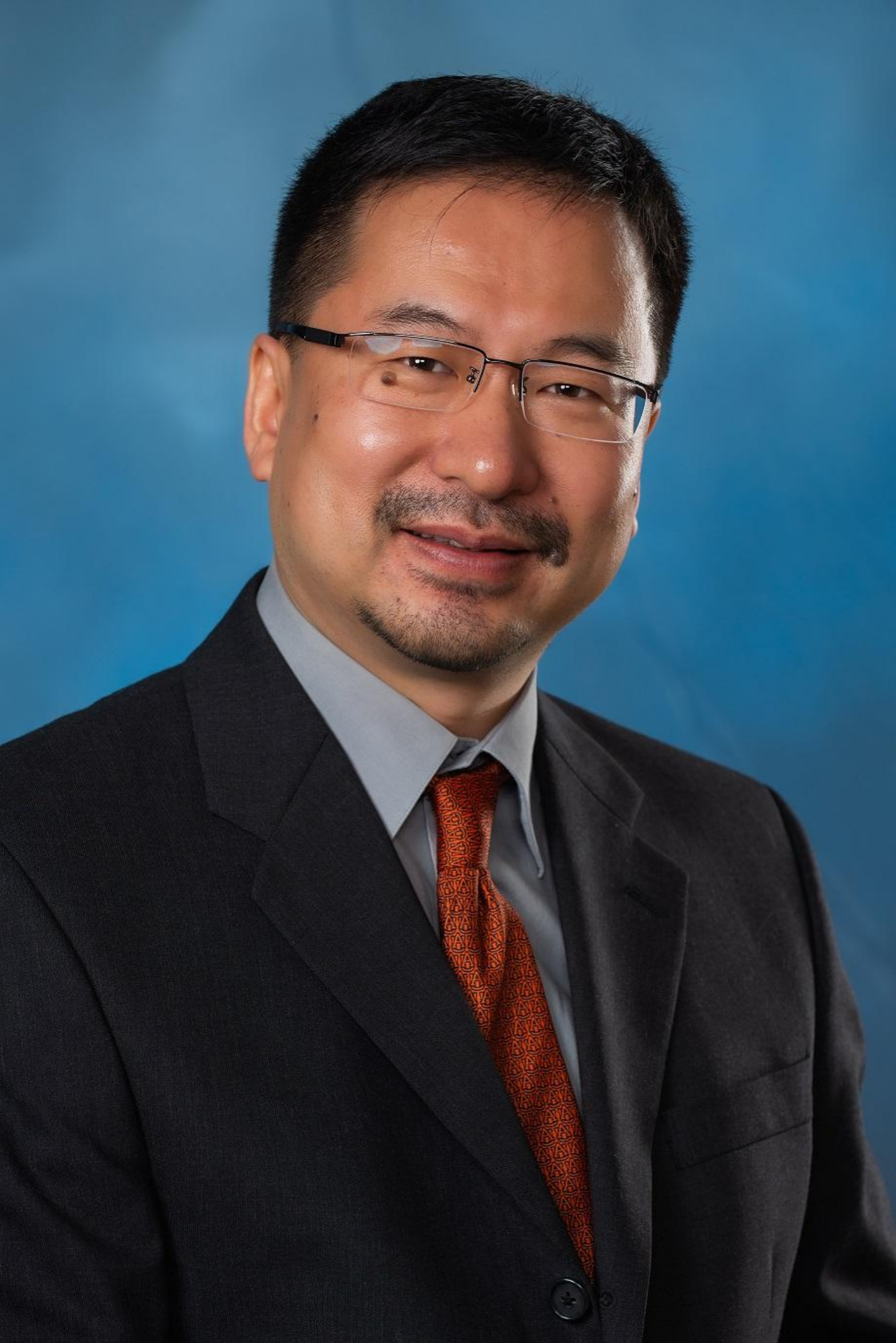}}]{Shiwen Mao}
	(S'99-M'04-SM'09-F'19) received a Ph.D in electrical and computer engineering from Polytechnic University, Brooklyn, N.Y. in 2004. He is the Samuel Ginn Professor and Director of Wireless Engineering Research and Education Center at Auburn University, Auburn, AL. His research interests include wireless networks, multimedia communications, and smart grid. He is a recipient of the IEEE ComSoc TC-CSR Distinguished Technical Achievement Award in 2019, the NSF CAREER Award in 2010, and the 2004 IEEE Communications Society Leonard G. Abraham Prize in the Field of Communications Systems. He is a Fellow of the IEEE.
\end{IEEEbiography}

\begin{IEEEbiography}[{\includegraphics[width=1in,height=1.25in,clip,keepaspectratio]{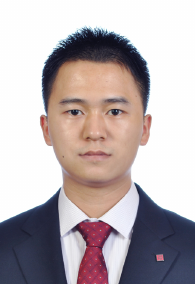}}]{Ruisi He}
	(Senior Member, IEEE) received the B.E. and Ph.D. degrees from Beijing Jiaotong University (BJTU), Beijing, China, in 2009 and 2015, respectively.
	
	Since 2015, Dr. He has been with the State Key Laboratory of Rail Traffic Control and Safety, BJTU, where he has been a Full Professor since 2019. Dr. He has been a Visiting Scholar in Georgia Institute of Technology, USA, University of Southern California, USA, and Universit\'e Catholique de Louvain, Belgium. His research interests include measurement and modeling of wireless channels, machine learning and clustering analysis in communications, vehicular and high-speed railway communications, 5G massive MIMO and high frequency communication techniques. He has authored/co-authored 3 books, 3 book chapters, more than 100 journal and conference papers, as well as several patents.
	
	Dr. He is an Editor of the IEEE Transactions on Wireless Communications, the IEEE Antennas and Propagation Magazine, and the IEEE Communications Letters. He serves as the Early Career Representative (ECR) of Commission C, International Union of Radio Science (URSI). He has been a Technical Program Committee (TPC) chair and member for many conferences and workshops. He received 2017-2019 Young Talent Sponsorship Program of China Association for Science and Technology, the Second Prize of the Natural Science Award for Scientific Research Achievements of the Ministry of Education in China in 2016, the Best Ph.D. Thesis Award of Chinese Institute of Electronics in 2016, the URSI Young Scientist Award in 2015, and five Best Paper Awards in conferences. He is a member of the European Cooperation in Science and Technology.
\end{IEEEbiography}

\begin{IEEEbiography}[{\includegraphics[width=1in,height=1.25in,clip,keepaspectratio]{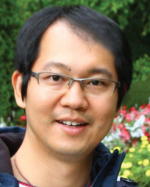}}]{Ning Wang}
	received the B.E. degree in communication engineering from Tianjin University, Tianjin, China, in 2004, the M.A.Sc. degree in electrical engineering from The University of British Columbia, Vancouver, BC, Canada, in 2010, and the Ph.D. degree in electrical engineering from the University of Victoria, Victoria, BC, Canada, in 2013. From 2004 to 2008, he was with the China Information Technology Design and Consulting Institute, as a Mobile Communication System Engineer, specializing in planning and design of commercial mobile communication networks, network traffic analysis, and radio network optimization. From 2013 to 2015, he was a Postdoctoral Research Fellow with the Department of Electrical and Computer Engineering, The University of British Columbia. Since 2015, he has been with the School of Information Engineering, Zhengzhou University, Zhengzhou, China, where he is currently an Associate Professor. He also holds adjunct appointments with the Department of Electrical and Computer Engineering, McMaster University, Hamilton, ON, Canada, and the Department of Electrical and Computer Engineering, University of Victoria, Victoria, BC, Canada. His research interests include resource allocation and security designs of future cellular networks, channel modeling for wireless communications, statistical signal processing, and cooperative wireless communications. He has served on the technical program committees of international conferences, including the IEEE GLOBECOM, IEEE ICC, IEEE WCNC, and CyberC. He was on the Finalist of the Governor Generals Gold Medal for Outstanding Graduating Doctoral Student with the University of Victoria in 2013.
\end{IEEEbiography}

\begin{IEEEbiography}[{\includegraphics[width=1in,height=1.25in,clip,keepaspectratio]{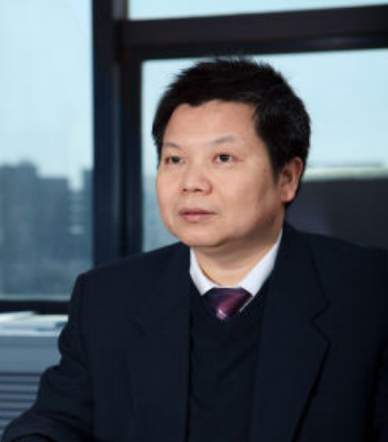}}]{Zhangdui Zhong}
	(SM'16-F'22) received the B.E. and M.S. degrees from Beijing Jiaotong University, Beijing, China, in 1983 and 1988, respectively.
	
	He is currently a Professor and an Advisor of Ph.D. candidates with Beijing Jiaotong University, where he is also currently a Chief Scientist of the State Key Laboratory of Rail Traffic Control and Safety. He is also the Director of the Innovative Research Team, Ministry of Education, Beijing, and a Chief Scientist of the Ministry of Railways, Beijing.
	
	He is also an Executive Council Member of the Radio Association of China, Beijing, and a Deputy Director of the Radio Association, Beijing. His interests include wireless communications for railways, control theory and techniques for railways, and GSM-R systems. His research has been widely used in railway engineering, such as the Qinghai-Xizang railway, DatongQinhuangdao Heavy Haul railway, and many high-speed railway lines in China. He has authored or co-authored seven books, five invention patents, and over 200 scientific research papers in his research area. Prof. Zhong was a recipient of the Mao YiSheng Scientific Award of China, Zhan TianYou Railway Honorary Award of China, and
	Top 10 Science/Technology Achievements Award of Chinese Universities.
	\vspace*{25mm}
\end{IEEEbiography}

\begin{IEEEbiography}[{\includegraphics[width=1in,height=1.25in,clip,keepaspectratio]{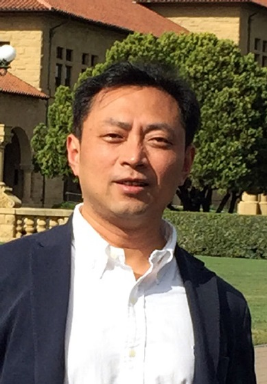}}]{Bo Ai}
	received the M.S. and Ph.D. degrees from Xidian University, China. He studies as a Post-Doctoral Student at Tsinghua University. He was a Visiting Professor with the Electrical Engineering Department, Stanford University, in 2015. He is currently with Beijing Jiaotong University as a Full Professor and a Ph.D. Candidate Advisor. He is the Deputy Director of the State Key Lab of Rail Traffic Control and Safety and the Deputy Director of the International Joint Research Center. He is one of the main people responsible for the Beijing Urban Rail Operation Control System, International Science and Technology Cooperation Base. He is also a Member, of the Innovative Engineering Based jointly granted by the Chinese Ministry of Education and the State Administration of Foreign Experts Affairs. He was honored with the Excellent Postdoctoral Research Fellow by Tsinghua University in 2007.
	
	He has authored/co-authored eight books and published over 300 academic research papers in his research area. He holds 26 invention patents. He has been the research team leader for 26 national projects. His interests include the research and applications of channel measurement and channel modeling, dedicated mobile communications for rail traffic systems. He has been notified by the Council of Canadian Academies that, based on Scopus database, he has been listed as one of the Top 1\% authors in his field all over the world. He has also been feature interviewed by the IET Electronics Letters. He has received some important scientific research prizes.
\end{IEEEbiography}

\end{document}